\documentclass[%
 reprint,
 amsmath,amssymb,
 aps,
floatfix,
]{revtex4-2}
\usepackage{graphicx}
\usepackage{hyperref}
\usepackage{dcolumn}
\usepackage{bm}
\usepackage[version=4]{mhchem}
\usepackage{enumerate}
\usepackage{enumitem, kantlipsum}
\newcommand{\RomanNumeralCaps}[1]{\MakeUppercase{\romannumeral #1}}

\begin{document}
\preprint{APS/123-QED}

\title{Incomplete fusion in $^{193}$Ir($^{12}$C, x)$^{205}$Bi reaction at $E_{lab}$ $\approx$ 5-7 AMeV}

\author{Amanjot$^1$}
\email{amanjot.19phz0004@iitrpr.ac.in, amuklair94@gmail.com}
\author{Priyanka$^1$}%
\author{Subham Kumar$^1$}%
\author{Rupinderjeet Kaur$^1$}%
\author{Malika Kaushik$^1$}%
\thanks{Presently at: Universit\`a degli Studi di Napoli Federico \RomanNumeralCaps{2}}
\author{Manoj Kumar Sharma$^2$}%
\author{Yashraj Jangid$^3$}
\author{Pushpendra P. Singh$^1$}%
\email{pps@iitrpr.ac.in}

\affiliation{$^1$Department of Physics, Indian Institute of Technology Ropar, Rupnagar - 140 001, Punjab, India}
\affiliation{$^2$Department of Physics, University of Lucknow, Lucknow - 226 007, Uttar Pradesh, India}
\affiliation{$^3$Inter-University Accelerator Centre, New Delhi - 110 067, India}

\date{\today}

\begin{abstract}
Low-energy heavy-ion induced reactions often involve incomplete fusion, but the dependence of ICF on various entrance-channel parameters remains unclear. In this work, we measure channel-by-channel production cross-sections of different evaporation residues populated via complete and/or incomplete fusion in $^{12}$C+$^{193}$Ir system at $E_{lab}$ $\approx$ 64--84 MeV ($\approx$ 5--7 AMeV) using the stacked-foil activation technique followed by offline $\gamma$-spectroscopy. Experimentally measured excitation functions have been analyzed in the framework of the statistical model code PACE4 using different values of the level-density parameter ($a$ = A/9-A/15 MeV${^{-1}}$). In the analysis of excitation functions, the $xn$ and $pxn$ channels (after correcting with their precursor contributions) have been explained fairly well with $a$ = A/13 MeV${^{-1}}$; however, almost all $\alpha$-emitting channels showed substantial enhancement over PACE4 predictions, which has been attributed to incomplete fusion. The incomplete fusion fraction ($F_{ICF}$) increases linearly with energy from 12\% to 18\% at 64 and 84 MeV, respectively. For better insights into the onset and strength of ICF, the variations of $F_{ICF}$ have been studied as a function of different entrance-channel parameters, which are found to increase with mass asymmetry, Coulomb factor, and neutron skin thickness. Further analysis of the data suggests the onset of ICF below the critical angular momentum ($\ell<\ell_{crit}$). Projectile breakup-driven incomplete fusion is found to suppress complete fusion by $\approx12\%$ and $\approx6\%$ w.r.t. the universal fusion function and the improved fusion function, respectively. These findings highlight the critical role of projectile structure at 5--7 AMeV energies, with implications for high-spin spectroscopy and reaction modeling.
\end{abstract}

\maketitle


\section{\label{sec:level1}Introduction}
Low-energy heavy-ion (HI) induced reactions are crucial for generating novel and intriguing nuclear states for spectroscopic investigations and facilitating the production of new nuclear species \cite{broda2006,hofmann2007,rudolph2013,van2019,*van2020,mougeot2021,dullmann2022,schneider1994}. Generally, nuclear reactions below 8 AMeV predominantly lead to the population of complete (CF) and incomplete fusion (ICF) residues. In the case of CF, for driving input angular momentum $\ell<\ell_{crit}$, the projectile completely coalesces with the target nucleus, forming an excited compound nucleus (CN) involving all nucleonic degrees of freedom of the interacting partners \cite{pochodzalla1986,wilschut1984}. However, the ICF occurs for relatively higher $\ell$-values ($\ell>\ell_{crit}$), in which the projectile dissociates into its constituents \cite{siwek1979,*siwek_1979,sum1980,*wilczynski1973}. One of the fragments fuses with the target nucleus to form an incompletely fused composite system, and the remnant is released in the forward cone, nearly undeflected, with the projectile velocity \cite{britt1961,galin1974,zolnowski1978}. Peripheral collisions have been reported to create favorable conditions for ICF reactions \cite{gerschel1982,beene1981,barker1980}. 

Several theoretical models, e.g., breakup fusion (BUF) model \cite{BUF1980}, sum-rule model \cite{sum1980}, promptly emitted particles (PEPs) model \cite{PEP1980}, etc., are known to elucidate ICF dynamics to some extent at energies $>$ 10 AMeV, in specific instances. However, the intricacies of ICF at energies as low as 3--5 AMeV remain poorly understood. Morgenstern et al. \cite{morgenstern1984,*morgenstern1982} correlated the onset of ICF with entrance-channel mass asymmetry ($\mu_{m}$) at energies $>$ 10 AMeV. However, at low incident energies, in ref. \cite{babu2003_ef,*babu2004_rrd,pps2008,*pps_spin2009,*pps2009,harish2017,*harish2019,*harish2015,*harish_thesis,kamal2014,*kamal2013,*kkamal2013,aquib2024,*aquib2025,dsingh_2018,*dsingh2018,*dsingh2017}, the projectile-dependence of $\mu_{m}$ systematics has been observed. The influence of projectile structure on ICF in terms of $Q_\alpha$ \cite{vijay2014,yadav2023,*yadav2017,*yadav_13C_2012,*yadav_12C_2012,tali2019,*tali2018,agarwal2022,*agarwal2021}, and neutron excess projectile \cite{vijay2014,agarwal2021,jashwal2023} has been realized. Sharma et al. \cite{vijay2014} found that $1n$ excess in $^{13}$C leads to a reduced ICF fraction ($F_{ICF}$) by $\sim$ 35--40\% due to a more negative $Q_\alpha$ value (-10.6 MeV) vs. -7.4 MeV for $^{12}$C projectile, suggesting neutron excess stabilizes breakup. The interplay between breakup, transfer, CF, and ICF near the Coulomb barrier has been clearly demonstrated by Gollan et al. \cite{gollan2021}, where breakup and $1n$ transfer significantly influence fusion dynamics. Rafiei et al. \cite{rafiei2010} indicated a non-dependence on $Z_{T}$, while Hinde et al. \cite{hinde2002}, and Gomes et al. \cite{gomes2004} suggested a near-proportional relationship with $Z_{T}$. Shuaib et al. \cite{shuaib2016} suggested a linear increase in ICF strength with the Coulomb factor ($Z_{P}$$Z_{T}$). However, upon re-examining the relationship between $Z_{P}$$Z_{T}$ and ICF, it has been found that ICF exhibits a linear trend separately for each projectile \cite{tali2018}. Further, $F_{ICF}$ in terms of target deformation has been discussed in refs. \cite{dsingh2018,munish2019,ojha2021,aquib2025}. Giri et al. \cite{giri2019} indicated that $F_{ICF}$ increases exponentially with deformation parameter ($\beta_2^T$), deformation length ($\beta_2^TR^T$), and neutron excess $(N-Z)^T$ rather than linear growth, indicating projectile structure dependence of ICF. The additional parameters such as $Z_{P}$$Z_{T}\beta_2^T$, $Z_{P}$$Z_{T}/(1-\beta_2^T)$ \cite{harish2019}, $\beta_2^T\mu_m$, $Z_{P}$$Z_{T}\mu_m$ \cite{dsingh_2018} have been introduced without much explanation of their relation with different entrance-channel parameters. These combined parameters are not suitable for explaining ICF, especially for spherical and deformed targets. Nasirov et al. \cite{nasirov2023} found that the increased fusion barrier, angular velocity, and decreased excitation energy contribute to the ICF at a large orbital angular momentum. Eudes et al. \cite{eudes2014} describe fusion excitation functions (EFs) independently in terms of $\mu_{m}$ and isospin. Yadav et al. \cite{yadav2023} elucidated the trends of ICF by considering both the total asymmetry and the system parameters, which appear to provide a more satisfactory explanation of the experimental data. ICF reactions have also been exploited to selectively populate high-spin states in heavy nuclei, as reported by Dejbakhsh et al. \cite{dejbakhsh1995}, which was noted by several other studies reporting the high-spin states in the final reaction products compared to CF \cite{xu2023,lee2020,dsingh2017,pps_spin2009,*pps2009}. Direct experimental evidence for the population of superdeformed states via ICF reactions was provided by Kaci et al. \cite{kaci1997}. The capability of ICF to populate complex nuclear configurations was further established by Dracoulis et al. \cite{dracoulis1998}, who identified single and multi-quasiparticle states. Recently, Burns et al. \cite{burns2025} have populated high-spin states of Erbium. Bossche et al. \cite{van2020} demonstrated that ICF at near-barrier energies can lead to the production of heavy and superheavy nuclei, emphasizing the sensitivity of ICF yields to entrance-channel dynamics and angular momentum.

This work probes ICF at 5--7 AMeV, revealing onset below $\ell<\ell_{crit}$ and projectile-dependent suppression unexplored in the $^{12}$C+$^{193}$Ir system. The channel-by-channel EFs of several evaporation residues (ERs) populated via $xn$, $pxn$, $\alpha xn$, and $2\alpha xn$ channels have been measured at energies 5--7 AMeV and analyzed in the framework of the statistical model code PACE4.

\section{\label{sec:level2}Experimental Methodology}
Experiments were conducted at the Inter-University Accelerator Centre (IUAC), New Delhi, using the stacked-foil activation technique followed by offline $\gamma$-spectroscopy. $^{193}$Ir targets of thickness 17--60 $\mu g/cm{^2}$ were fabricated on 1--1.5 $mg/cm{^2}$ Al foils \cite{amanjot2024}. The stacks were irradiated in the General Purpose Scattering Chamber (GPSC) \cite{gpsc} with $^{12}$C$^{6+}$ beams at 81 and 84 MeV, with 1--2 pnA beam current, for 6--8 hours. The Al backing served both as a catcher for ERs and as an energy degrader (about 1--2 MeV per foil) for the subsequent target foils in the stack \cite{srim}. After irradiations, the stacks were counted offline using pre-calibrated high-resolution clover HPGe detectors coupled with a CAMAC-based DAQ system \cite{subramaniam2006, candle}. Standard $\gamma$-sources ($^{152}$Eu, $^{60}$Co, and $^{133}$Ba) were used for energy and efficiency calibration.

\begin{figure}[]
\centering
\includegraphics[trim=6.5cm 0.3cm 1.5cm 0.3cm, width=9.5cm]{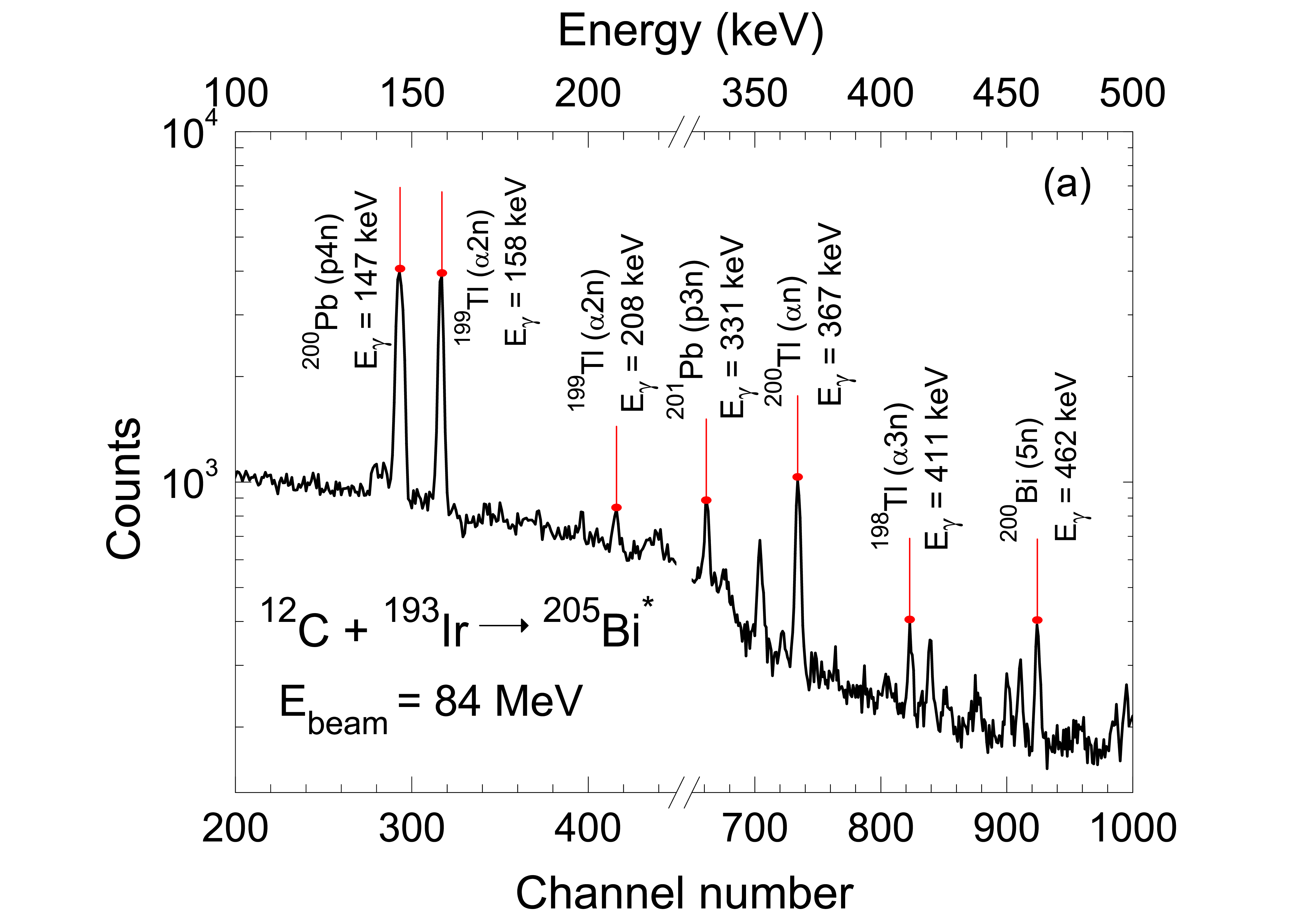}
\includegraphics[trim=2.5cm 0.3cm 3.5cm 0.3cm, width=8.5cm]{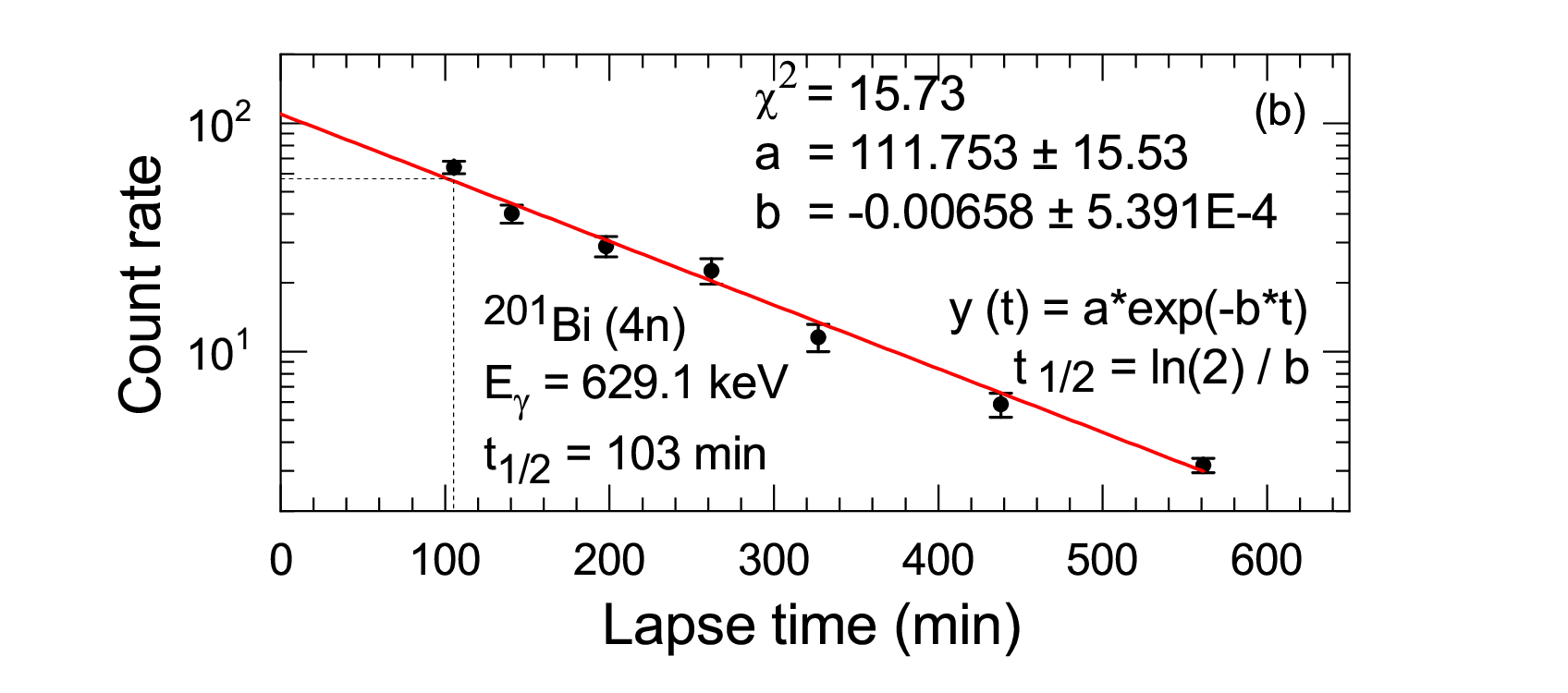}
\caption{$(a)$ A portion of $\gamma$-ray spectrum obtained at 84 MeV for $^{12}$C+$^{193}$Ir system, $(b)$ the decay curve of $^{201}$Bi nuclei reproducing a half-life of 103 min by following $E_\gamma$ = 629.1 keV.}
\label{fig1}
\end{figure}

\begin{table}
\caption{List of identified ERs populated in $^{12}$C+$^{193}$Ir system along with their spectroscopic properties.}
\label{tab:table1}
\begin{tabular*}{\columnwidth}{@{\extracolsep{\fill}}llllll@{}}
\hline
Channel & ER & Half-life  & Spin  &  Gamma-line   &  Intensity \\
&  & $t_{1/2}$  & $J^\pi$  & $E_\gamma$ (keV)  & $I_\gamma$ ($\%$) \\
\hline
$4n$ & $^{201}$Bi & 103 min & 9/2$^{-}$ & 629.10 & 26.0 \\ 
$5n$ & $^{200}$Bi & 36.4 min & 7$^{+}$ & 462.34 & 98.0 \\ 
$p3n$ & $^{201}$Pb & 9.33 h & 5/2$^{-}$ & 331.15 & 77.0 \\ 
$p4n$ & $^{200}$Pb & 21.5 h  & 0$^{+}$ & 147.63 & 38.2 \\ 
$\alpha n$ & $^{200}$Tl & 26.1 h  & 2$^{-}$ & 367.94 & 87.0 \\ 
$\alpha2n$ & $^{199}$Tl & 7.42 h  & 1/2$^{+}$ & 208.20 & 12.3 \\ 
$\alpha3n$ & $^{198}$Tl$^g$ & 5.3 h  & 2$^{-}$ & 411.80 & 80.0 \\ 
$2\alpha n$ & $^{196}$Au$^m$ & 9.6 h  & 12$^{-}$ & 147.81 & 43.5 \\ 
\hline
\end{tabular*}
\end{table}

The ERs have been identified by their characteristic $\gamma$-rays and confirmed through decay curve analysis. Fig.~\ref{fig1}$(a)$ shows a portion of the $\gamma$-ray spectrum obtained at 84 MeV in which the peaks of different ERs are marked. As a representative case, the decay curve of $^{201}$Bi(4$n$) given in Fig.~\ref{fig1}$(b)$, obtained by following the $\gamma$-line $E_{\gamma}$ = 629.1 keV \cite{nndc}, has been analyzed using a $\chi^2$-minimized exponential fit yielding a half-life of 105 min close to the literature value of 103 min. The same approach has been followed to identify all ERs listed in Table~\ref{tab:table1}. The production cross-sections of ERs calculated using a standard formulation \cite{prasad2018} are listed in Tables~\ref{tab:table2} and ~\ref{tab:table3}. The overall uncertainty ($\leq$13$\%$) in the measured cross-sections incorporate quadratic summation of target thickness ($\sim$2\%), beam flux ($\sim$5\%), detection efficiency ($\sim$5\%), counting statistics ($\sim$1--3\%) and dead time corrections ($\leq$1\%), excluding the uncertainties in nuclear data \cite{browne,nuclearwallet}.

\begin{table*}
\caption{Experimental cross-sections ($\sigma_{exp}$) of ERs alongwith PACE4 calculations ($\sigma_{PACE}$) for $K$ = 13 populated in the $^{12}$C+$^{193}$Ir system via CF process.}
\label{tab:table2}
\begin{tabular*}{\textwidth}{@{\extracolsep{\fill}}lccccccccl@{}}
\hline
$E_{lab}$ & $E_{c.m.}$  & $^{201}$Bi ($4n$) & $^{201}$Bi & $^{200}$Bi ($5n$) & $^{200}$Bi & $^{201}$Pb ($p3n$) & $^{201}$Pb & $^{200}$Pb ($p4n$) & $^{200}$Pb \\
(MeV) & (MeV) & $\sigma_{exp}$  & $\sigma_{PACE}$  & $\sigma_{exp}$ & $\sigma_{PACE}$ & $\sigma_{exp}$  & $\sigma_{PACE}$  &$\sigma_{exp}$ & $\sigma_{PACE}$ \\ 
&  &  (mb) & (mb) & (mb) & (mb) & (mb) & (mb) & (mb) &  (mb)\\  \hline
83.9 $\pm$ 0.013 & 79.1 & 26.0 $\pm$ 6.7 & 25.6 & 626.4 $\pm$ 65.6 & 628 & 9.4 $\pm$ 3.1 & 9.57 & 83.8 $\pm$ 24.6 & 83.1 \\
80.9 $\pm$ 0.007 & 76.2  & 69.9 $\pm$ 16.8 & 61.8 & 693.2 $\pm$ 83.0 & 697 & 17.6 $\pm$ 5.5 & 17.5 & 67.0 $\pm$ 20.1 & 68.3 \\
79.4 $\pm$ 0.012 & 74.7  & 96.7 $\pm$ 15.6 & 96 & 662.9 $\pm$ 68.8 & 662 & 22.6 $\pm$ 5.4 &22.7 & 56.7 $\pm$ 14.5 & 55.8\\ 
76.5 $\pm$ 0.009  & 72.1  & 176.1 $\pm$ 23.3 & 173 & 562.2 $\pm$ 77.8 & 531 & 29.8 $\pm$ 6.6 & 29 & 32.0 $\pm$ 8.2 & 31.9\\ 
74.8 $\pm$ 0.026 & 70.4  & 250.9 $\pm$ 32.9 & 253 & 405.7 $\pm$ 47.8 & 402 & 32.5 $\pm$ 7.4 & 33.1 & 21.4 $\pm$ 7.2 & 19.4\\ 
71.9 $\pm$ 0.022 & 67.7 & 389.8 $\pm$ 40.5 & 387 & 182.5 $\pm$ 20.6 & 181 & 34.0 $\pm$ 7.4 & 35 & 7.4 $\pm$ 2.2 & 5.47\\ 
70.1 $\pm$ 0.018 & 66.0 & 401.1 $\pm$ 56.1 & 415 & 99.7 $\pm$ 14.4 & 96.8 & 29.9 $\pm$ 6.6 & 28 & 2.7 $\pm$ 0.7 & 2.29\\ 
64.4 $\pm$ 0.018 & 60.6  & 253.2 $\pm$ 25.6 & 252 & --& 0.08 & 6.8 $\pm$ 1.5 & 8.59 &  -- & 0.003\\ 
\hline
\end{tabular*}
\end{table*}

\begin{table*}[]
\caption{Experimental cross-sections ($\sigma_{exp}$) of ERs alongwith PACE4 calculations ($\sigma_{PACE}$) for $K$ = 13 populated in the $^{12}$C+$^{193}$Ir system via CF and/or ICF processes.}
\label{tab:table3}
\begin{tabular*}{\textwidth}{@{\extracolsep{\fill}}lccccccccl@{}}
\hline
$E_{lab}$ &  $E_{c.m.}$ & $^{200}$Tl ($\alpha n$) & $^{200}$Tl & $^{199}$Tl ($\alpha2n$) & $^{199}$Tl & $^{198}$Tl$^{g}$ ($\alpha3n$) & $^{198}$Tl &  $^{196}$Au$^m$ ($2\alpha n$) & $^{196}$Au\\
(MeV) & (MeV) & $\sigma_{exp}$  & $\sigma_{PACE}$  & $\sigma_{exp}$ & $\sigma_{PACE}$ & $\sigma_{exp}$  & $\sigma_{PACE}$  &$\sigma_{exp}$ & $\sigma_{PACE}$ \\ 
&  &  (mb) & (mb) & (mb) & (mb) & (mb) & (mb) & (mb) &  (mb)\\  \hline
83.9 $\pm$ 0.013 & 79.1 & 50.0 $\pm$ 9.4 & 0.496 & 26.3 $\pm$ 6.1 & 0.0539 & 18.5 $\pm$ 2.2 & 9.02 & 125.5 $\pm$ 20.2 & --\\ 
80.9 $\pm$ 0.007 & 76.2 & 112.3 $\pm$ 24.7 & 0.337 & 22.8 $\pm$ 6.6 & 0.159 & 33.5 $\pm$ 3.9 & 15.5 & 73.5 $\pm$ 14.0 & -- \\
79.4 $\pm$ 0.012 & 74.7 & 80.0 $\pm$ 13.3 & 0.226 & 22.3 $\pm$ 5.4 & 0.226 & 33.6 $\pm$ 3.9 & 19 & 83.6 $\pm$ 13.3 & --\\
76.5 $\pm$ 0.009 & 72.1 & 52.9 $\pm$ 15.9 & 0.0591 & 29.9 $\pm$ 5.0 & 0.616 & 43.7 $\pm$ 5.2 & 21.9 & 70.5 $\pm$ 9.8 & 0.0084\\
74.8 $\pm$ 0.026 & 70.4 & 28.9 $\pm$ 5.6 & 0.0313 & 27.8 $\pm$ 3.4 & 0.891 & 46.8 $\pm$ 4.9 & 23.6 & 78.1 $\pm$ 13.8 & 0.0469 \\
71.9 $\pm$ 0.022 & 67.7 & 35.0 $\pm$ 7.2 & 0.0133 & 27.7 $\pm$ 4.5 & 1.48 & 38.9 $\pm$ 5.5 & 20.2 & 41.6 $\pm$ 6.2 & 0.0133\\
70.1 $\pm$ 0.018 & 66.0 & 33.8 $\pm$ 7.5 & -- & 28.6 $\pm$ 4.9 & 2.26 & 31.0 $\pm$ 4.0 & 14.6 & 20.7 $\pm$ 3.7 & 0.0059\\
64.4 $\pm$ 0.018 & 60.6 & 11.4 $\pm$ 2.8 & 0.0155 & 20.9 $\pm$ 4.8 & 3.86 & 10.8 $\pm$ 1.4 & 2.93 & 8.3 $\pm$ 1.6 & 0.0062\\
\hline
\end{tabular*}
\end{table*}

\section{\label{sec:level3}Results and Analysis}
Experimentally measured cross-sections ($\sigma_{exp}$) of ERs, with overall errors, have been plotted against projectile energy to generate channel-by-channel EFs and analyzed using the statistical model code PACE4 \cite{gavron1980,lise}. In this code, the level density parameter, $a$ = A/K MeV${^{-1}}$, is crucial, where $A$ is the mass number of CN, and $K$ is a free parameter that may be adjusted to reproduce experimental EFs. Since the code describes only CF and subsequent statistical decay of CN, it does not include projectile breakup and partial capture mechanisms that generate ICF, hence, any discrepancies between experimental cross-sections and PACE4 predictions may indicate contributions from ICF in $\alpha$-emitting channels and/or pre-equilibrium emission processes, especially for neutron-emitting channels at higher energies \cite{manoj2016,*manoj2006,*manoj2003}. Detailed channel-by-channel analysis of EFs using PACE4 is provided in the following sections.

\subsection{$xn$ -- channels}
The CF mechanism proceeds via the formation of CN, followed by neutron evaporation as,
\begin{quote}
    $^{12}$C + $^{193}$Ir $\xrightarrow{}$ $^{205}$Bi$^*$ $\xrightarrow{xn}$ $^{205-x}$Bi; $x$ = 4, 5.
\end{quote}

\begin{figure}[]
\centering
\includegraphics[trim=1.0cm 0.3cm 1.5cm 0.3cm, width=8cm]{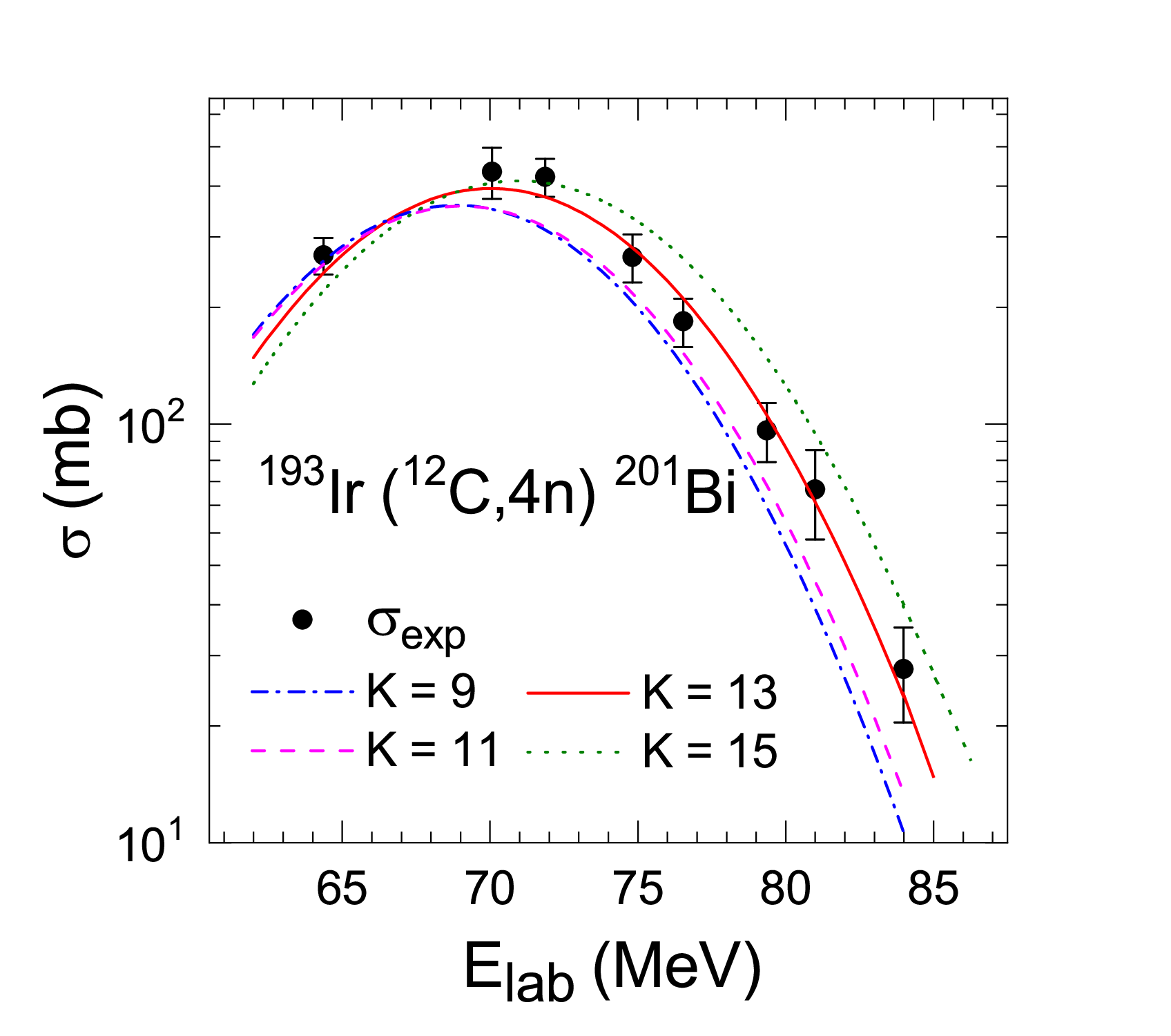}
\caption{Experimental EF of $^{201}$Bi($4n$) compared with PACE4 calculations using different values of $K$ = 9, 11, 13 and 15.}
\label{fig2}
\end{figure}

For an insight into how well the production of these residues can be explained by the formation and decay of CN, the EF of $^{201}$Bi($4n$) is compared with PACE4 using $K$ = 9, 11, 13, and 15 in Fig.~\ref{fig2}. As can be seen from this figure, the cross-sections of the $4n$ channel are reasonably reproduced by PACE4 using $a$ = A/13 MeV${^{-1}}$ within the experimental uncertainties, indicating that the population of $^{201}$Bi occurs only via CF. Moreover, for the $4n$ channel, both $\chi^2$ (0.325) and RMS deviation (4.36\%) are minimum for $K$ = 13, confirming its choice as an optimized parameter for analyzing other ERs formed in this system at the studied energy range. 

\begin{figure}[]
\centering
\includegraphics[trim=1.0cm 0.3cm 1.5cm 0.3cm, width=8cm]{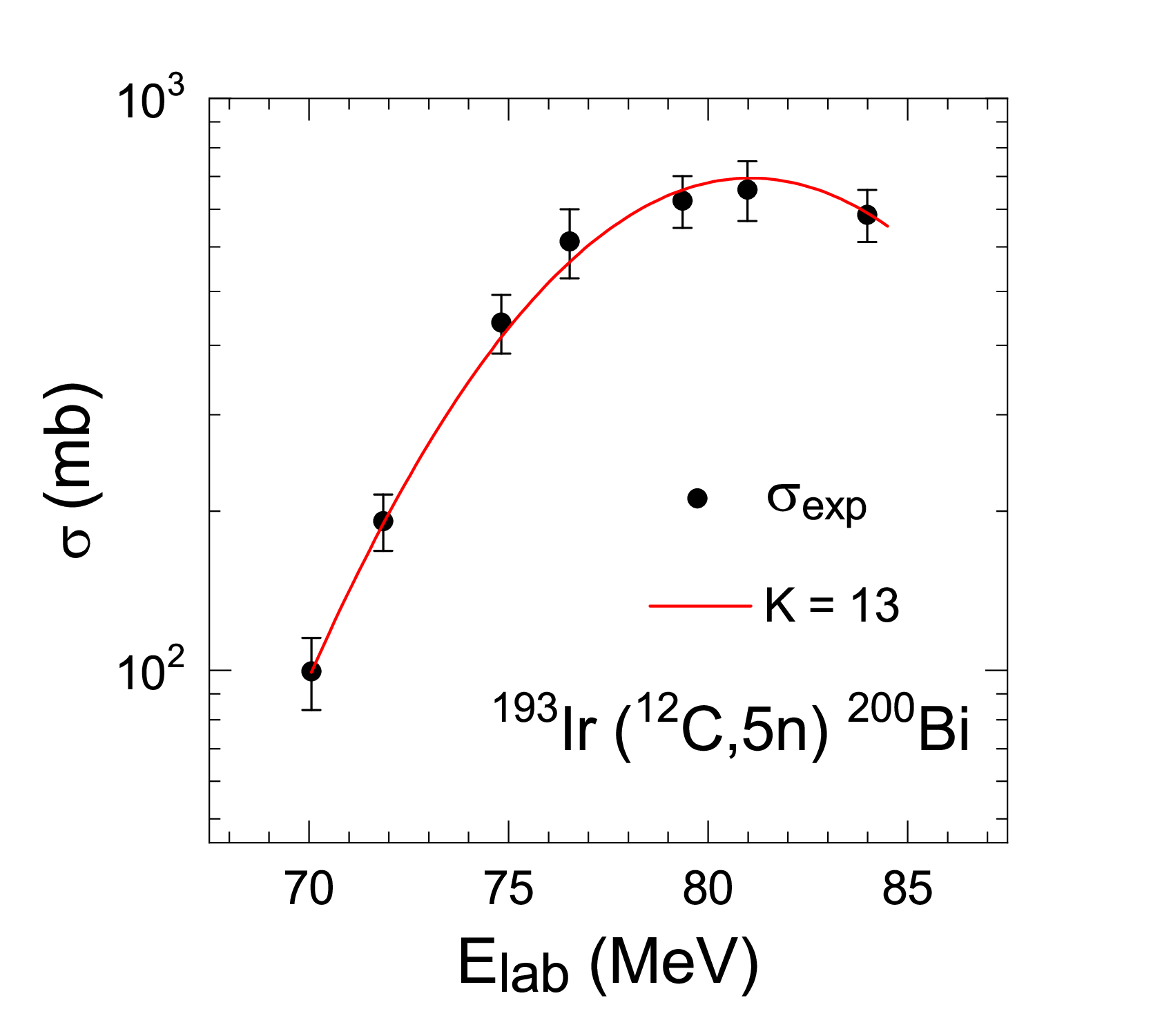}
\caption{The EF of $^{200}$Bi($5n$) with PACE4 for $K$ = 13.}
\label{fig3}
\end{figure}

The EF of $^{200}$Bi($5n$) plotted in Fig.~\ref{fig3} reproduces the experimental data fairly well for $K$ = 13. Based on results presented in Fig.~\ref{fig2} and ~\ref{fig3}, it can be inferred that the residues $^{201}$Bi and $^{200}$Bi are populated via emission of $4n$ and $5n$ channels, respectively, from a fully equilibrated CN $^{205}$Bi$^*$ formed via CF in $^{12}$C+$^{193}$Ir reaction.

\subsection{$pxn$ -- channels}
\begin{figure}[]
\centering
\includegraphics[trim=1cm 0.3cm 2.0cm 0.3cm, width=8.5cm]{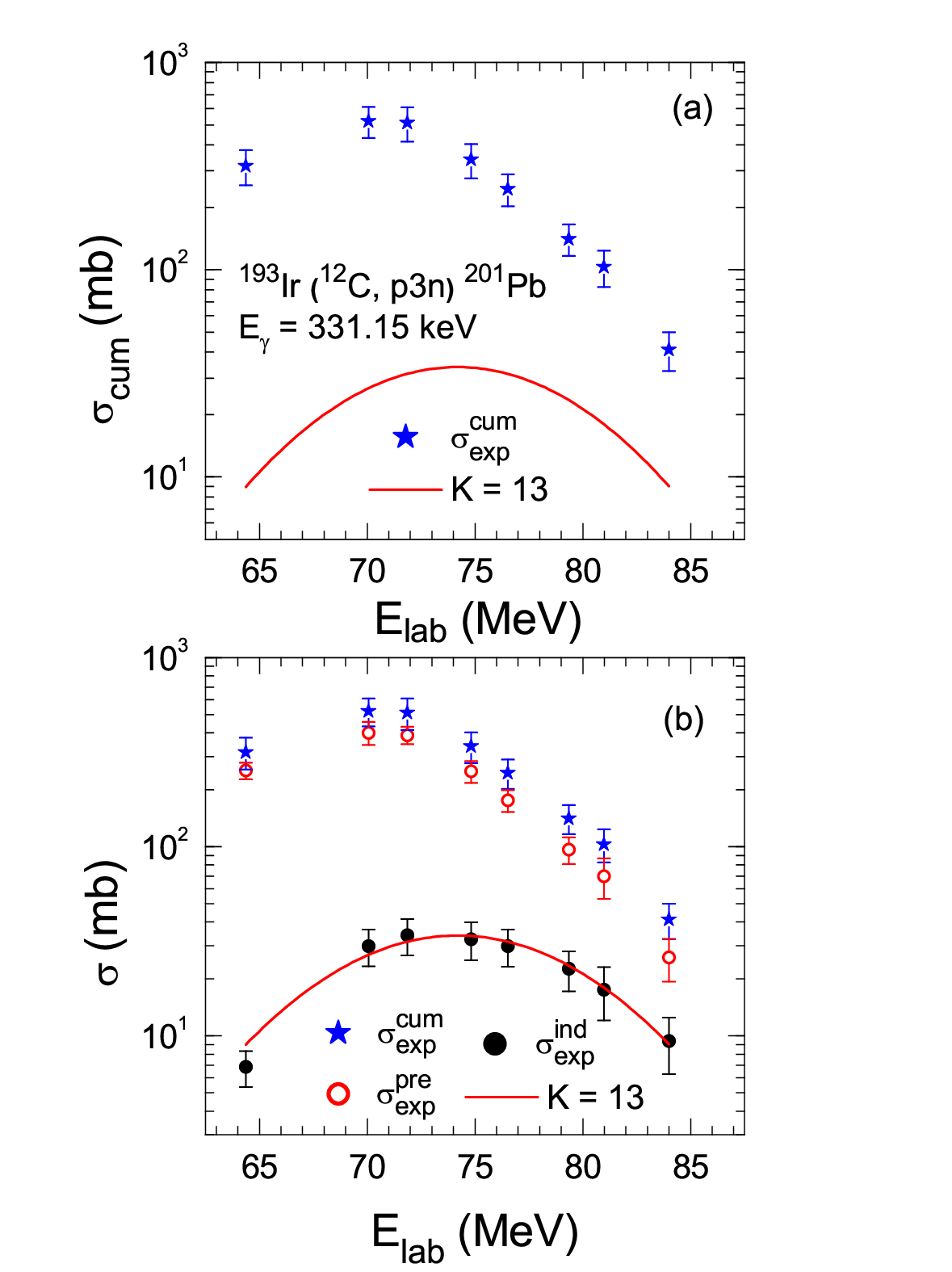}
\caption{$(a)$ The experimental EF of $^{201}$Pb (p3n) residue, $(b)$ $\sigma_{exp}^{cum}$, $\sigma_{exp}^{pre}$ and $\sigma_{exp}^{ind}$ contributions are compared with PACE4 calculations for $K$ = 13.}
\label{fig4}
\end{figure}

Fig.~\ref{fig4}$(a)$ shows the EF of $^{201}$Pb ($t_{1/2}$ = 9.33 h) populated via $p3n$ channel. However, PACE4 substantially under-predicts the EF, indicating the presence of its higher charge precursor, $^{201}$Bi, through $\beta^+$ emission and/or electron capture (EC) as,
\begin{quote}
$^{12}$C + $^{193}$Ir $\xrightarrow{}$ $^{205}$Bi$^*$ $\xrightarrow{xn}$ $^{205-x}$Bi (precursor nucleus); $x$ = 4 \\
$^{205}$Bi$^*$ $\xrightarrow{}$ $^{204-x'}$Pb + $p$ + $x'n$; $x'$ = 3\\
$^{205-x}$Bi $\xrightarrow{\beta^+ / EC}$ $^{204-x'}$Pb (daughter nucleus); \\
$x$ = 4 and $x'$ = 3, provided, $t_{1/2}^p$ $\ll$ $t_{1/2}^d$.
\end{quote}

In Fig.~\ref{fig4}$(b)$, the independent contribution of $^{201}$Pb has been obtained using $\sigma_{exp}^{ind} (mb) = \sigma_{exp}^{cum} - {F_P} \sigma_{exp}^{pre}$, where $\sigma_{exp}^{ind}$, $\sigma_{exp}^{cum}$, and $\sigma_{exp}^{pre}$ are the independent yield, cumulative yield, and precursor cross-section. The value of numerical coefficient $F_P = \frac{P_{pre} \lambda_{pre}} {\lambda_{pre}-\lambda_{d}}$, where $P_{pre}$ is the branching ratio of precursor decay to the daughter nucleus, $\lambda_{pre}$ and $\lambda_{d}$ are the decay constants of precursor and daughter nuclei, respectively, which is a successive radioactive decay formulation proposed by Cavinato et al. \cite{cavinato1995}. The independent cross-section of $^{201}$Pb($p3n$) is reproduced reasonably well by PACE4 for $K$ = 13 after subtracting the contribution of $^{201}$Bi from the cumulative cross-section. 

\begin{figure}[]
\centering
\includegraphics[trim=1.0cm 0.3cm 1.5cm 0.3cm, width=8cm]{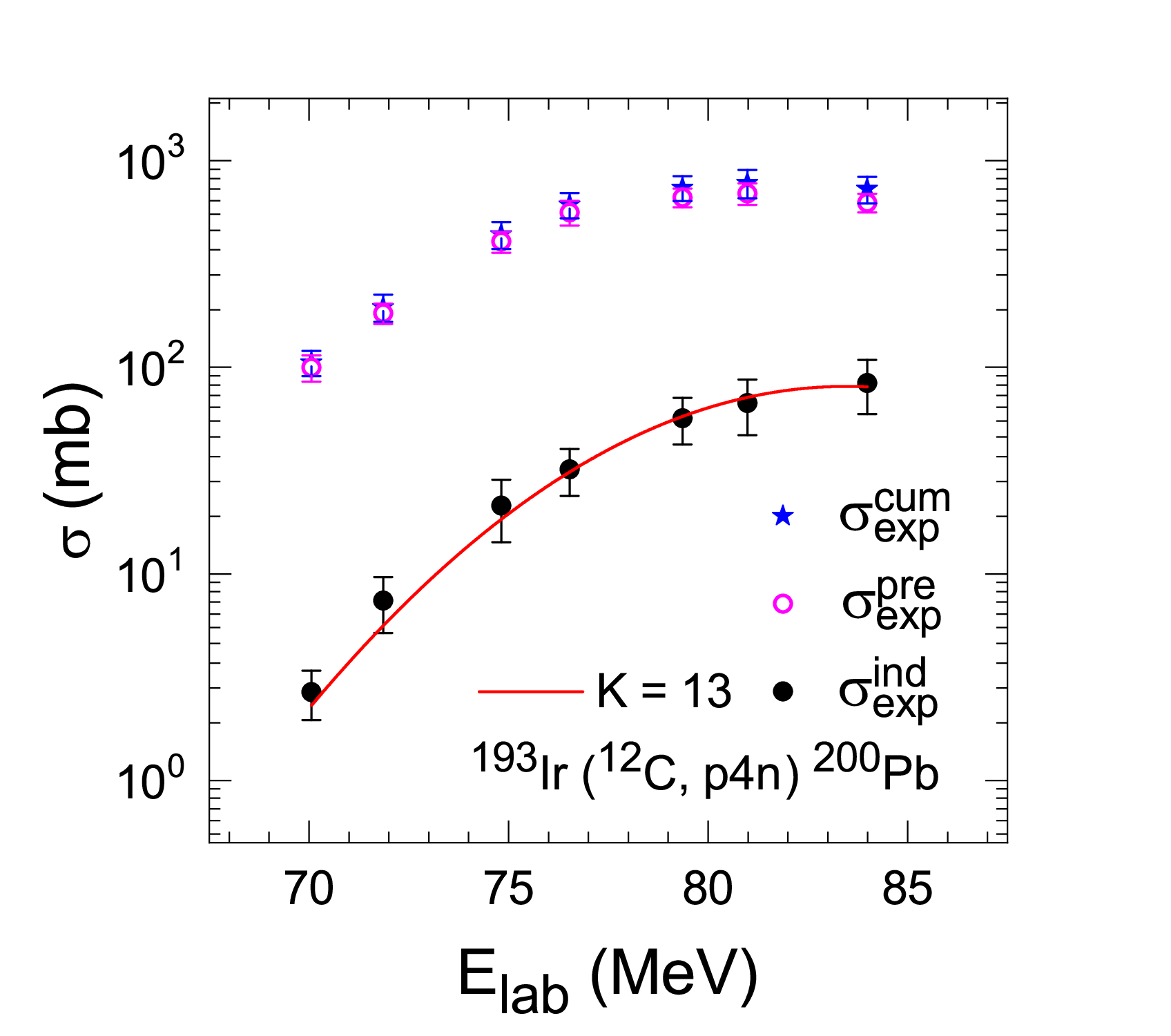}
\caption{Experimental EF of $^{200}$Pb(p4n) residue. The contributions of $\sigma_{exp}^{pre}$ and $\sigma_{exp}^{ind}$ are plotted with $\sigma_{exp}^{cum}$ and PACE4 calculations for $K$ = 13.}
\label{fig5}
\end{figure}

Similarly, the $^{200}$Pb($p4n$) residue ($t_{1/2}$ = 21.5 h) is fed by its higher-charge isobar, $^{200}$Bi($5n$), via $\beta^+$ and/or EC decay. In Fig.~\ref{fig5}, the value of $\sigma_{exp}^{ind}$ for $^{200}$Pb($p4n$) residue is compared with PACE4 along with $\sigma_{exp}^{pre}$ and $\sigma_{exp}^{cum}$. The value of $\sigma_{exp}^{ind}$ is in good agreement with the PACE4 for $a$ = A/13 MeV${^{-1}}$. Based on the results presented in Fig.~\ref{fig4}$(b)$ and ~\ref{fig5}, it can be inferred that the $^{201}$Pb and $^{200}$Pb are populated via CF of $^{12}$C with $^{193}$Ir.

\subsection{$\alpha xn$ and $2\alpha n$ -- channels}
The $\alpha$-emitting residues can be expected to be populated through both CF and/or ICF reaction as,
\begin{enumerate}
    \item  CF of $^{12}$C with $^{193}$Ir:\\ 
    $^{12}$C + $^{193}$Ir $\xrightarrow{}$ $^{205}$Bi$^*$ $\xrightarrow{}$ $^{201-x}$Tl + $\alpha$$xn$; ($x$ = 1)   
    \item  ICF of $^{12}$C with $^{193}$Ir:\\
     $^{12}$C ($^{8}$Be + $\alpha$) $\xrightarrow{}$ $^{8}$Be + $^{193}$Ir $\xrightarrow{}$ $^{201}$Tl$^*$ \\    
     $^{201}$Tl$^*$ $\xrightarrow{}$ $^{201-x}$Tl + $xn$ + $\alpha$;
($\alpha$: spectator, $x$ = 1).
\end{enumerate}

\begin{figure}[]
\centering
\includegraphics[trim=1cm 0.3cm 2.0cm 0.3cm, width=8.4cm]{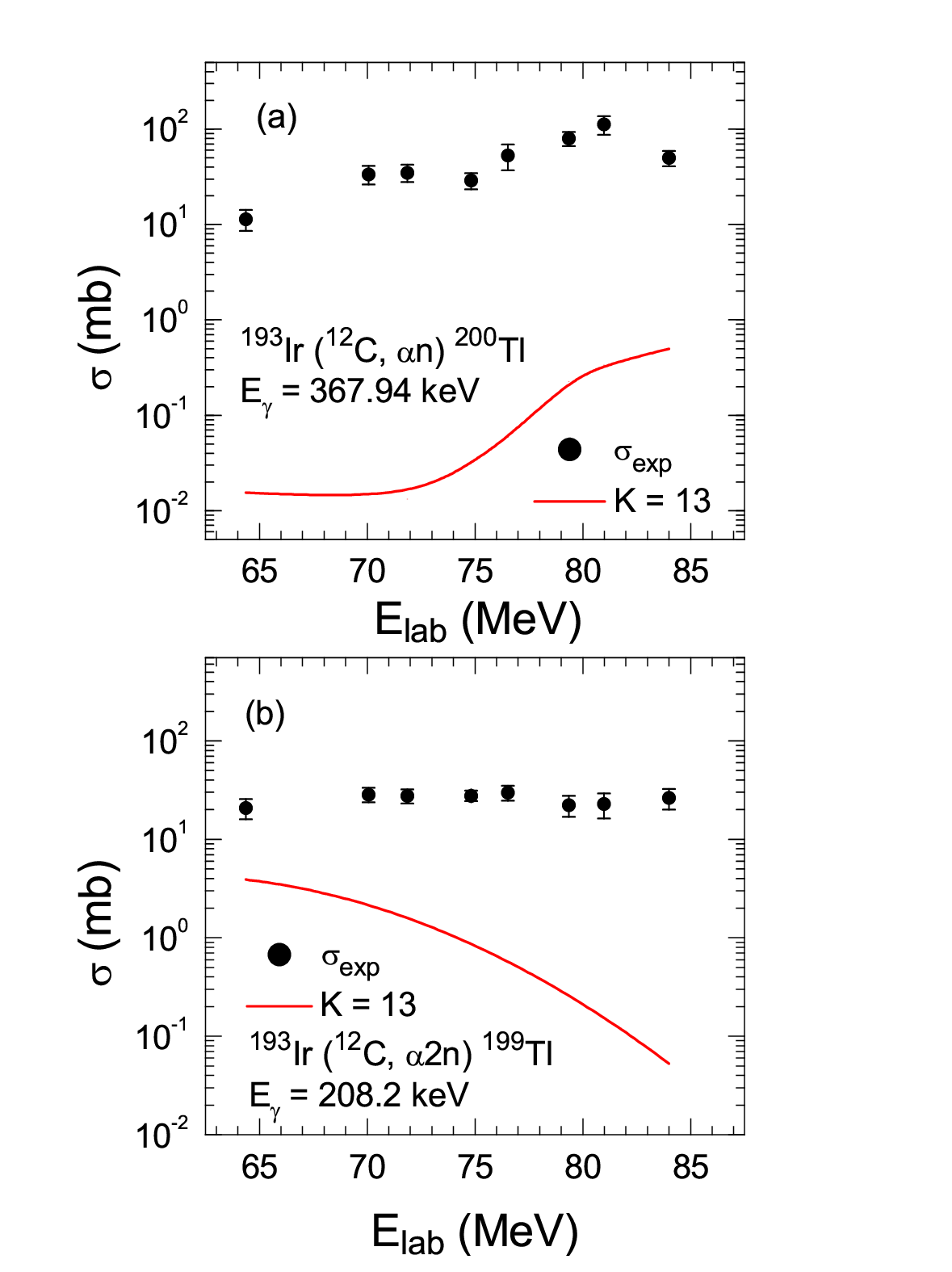}
\caption{Experimental EFs of $(a)$ $^{200}$Tl($\alpha n$), and $(b)$ $^{199}$Tl($\alpha2n$) expected to be populated via CF and/or ICF of $^{12}$C with $^{193}$Ir.}
\label{fig6}
\end{figure}

\begin{figure}[]
\centering
\includegraphics[trim=1cm 0.3cm 2.0cm 0.3cm, width=8.4cm]{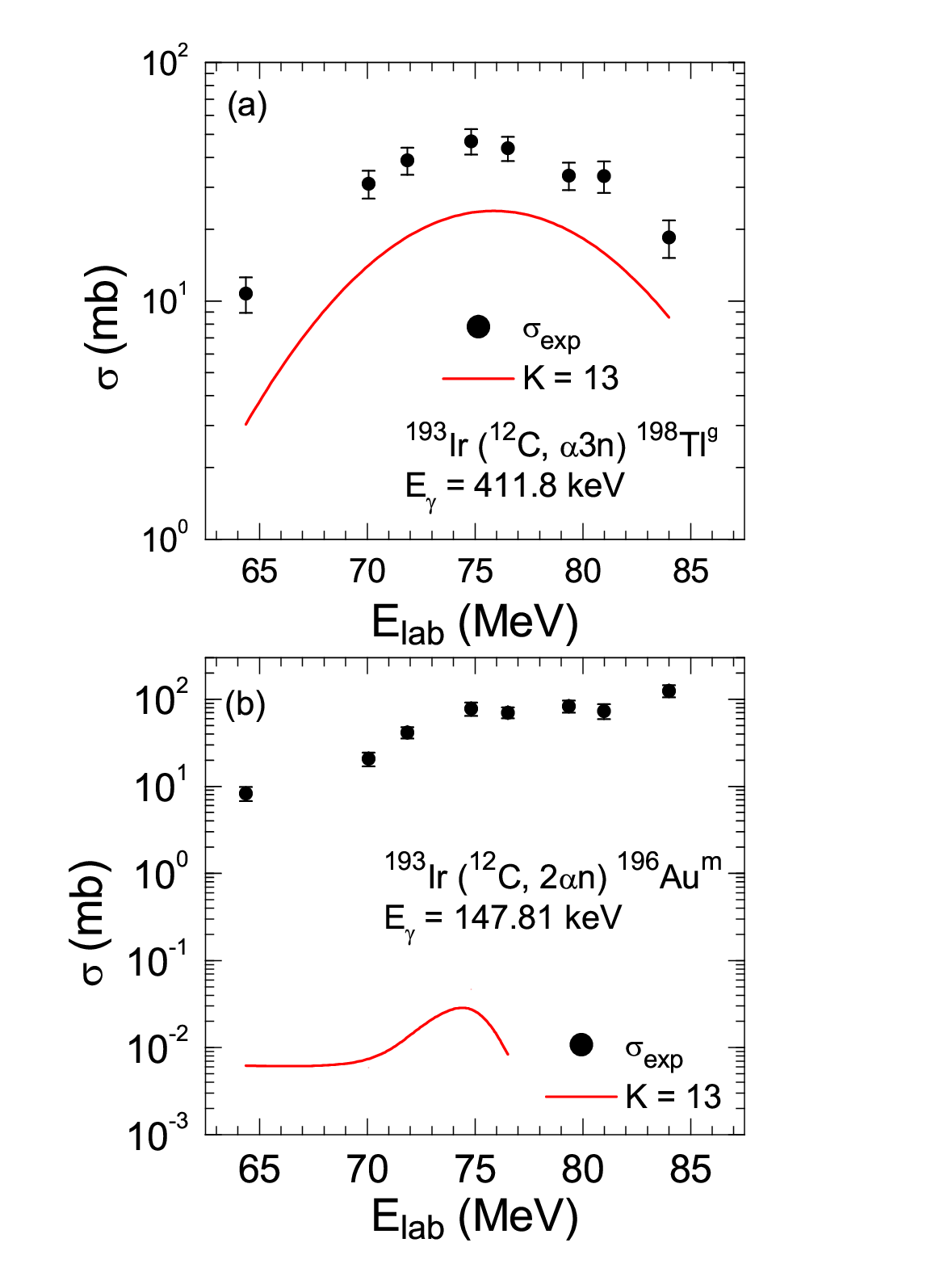}
\caption{Experimental EFs of $(a)$ $^{198}$Tl($\alpha3n$) in the ground state, and $(b)$ $^{196}$Au$^m$($2\alpha n$) populated in the metastable state compared with PACE4 calculations.}
\label{fig7}
\end{figure}

The EF of $^{200}$Tl($\alpha n$) residue shown in Fig.~\ref{fig6}$(a)$ exhibits a clear enhancement over the PACE4 calculations for $K$ = 13. Similarly, the experimental cross-sections for $^{199}$Tl($\alpha2n$) and $^{198}$Tl($\alpha3n$) residues are enhanced over the theoretical calculations as observed in Fig.~\ref{fig6}$(b)$ and ~\ref{fig7}$(a)$. Since PACE4 accounts only for CN decay following CF, this enhancement indicates the presence of an additional reaction mechanism, suggesting their population via ICF of $^{12}$C with $^{193}$Ir. The $^{196}$Au is populated in a metastable state, and an enhancement is clearly observed compared to PACE4 predictions as plotted in Fig.~\ref{fig7}$(b)$. This residue can be populated in two ways.
\begin{enumerate}
    \item  $2\alpha$ -- fusion:\\
    $^{12}$C ($^{8}$Be + $\alpha$) $\xrightarrow{}$ $^{8}$Be + $^{193}$Ir $\xrightarrow{}$ $^{201}$Tl$^*$ \\    
    $^{201}$Tl$^*$ $\xrightarrow{}$ $^{196}$Au + $n$ + $\alpha$;\\
    $\alpha$ particle is a spectator.
    \item  ${\alpha}$ -- fusion:\\
    $^{12}$C ($^{8}$Be + $\alpha$) $\xrightarrow{}$ ${\alpha}$ + $^{193}$Ir $\xrightarrow{}$ $^{197}$Au$^*$ \\    
    $^{197}$Au$^*$ $\xrightarrow{}$ $^{196}$Au + $n$;\\
    $2\alpha$ particle is a spectator.
\end{enumerate}

The significant contribution to the production of $^{196}$Au residue comes from the $2\alpha$, i.e., $^{8}$Be fusion.

\begin{figure}[]
\centering
\includegraphics[trim=1.0cm 0.3cm 1.5cm 0.3cm, width=8cm]{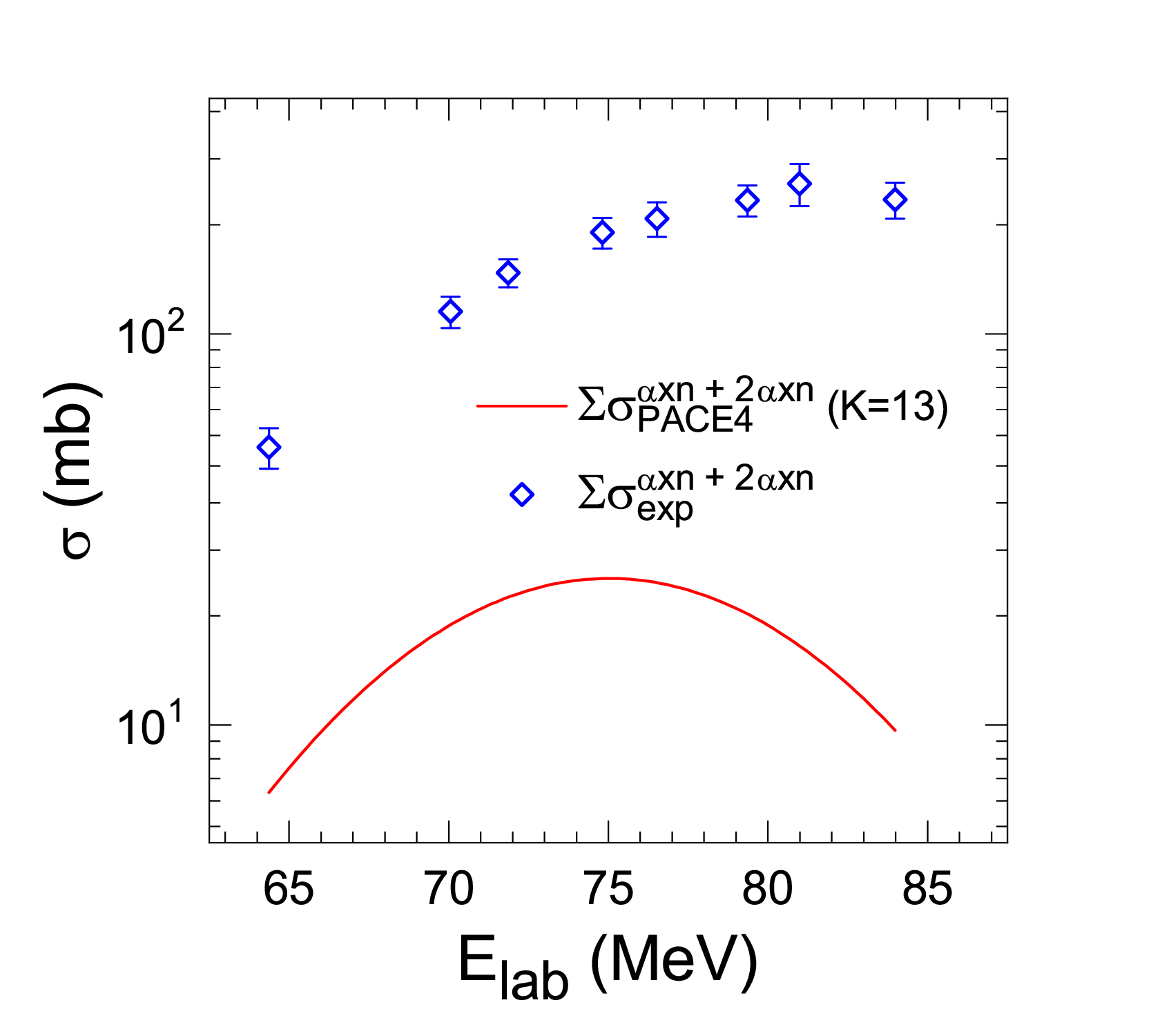}
\caption{Experimental EFs of all $\alpha$-emitting channels ($\Sigma$$\sigma_{exp}^{\alpha xn+2\alpha n}$) identified in $^{12}$C+$^{193}$Ir system compared with PACE4 ($\Sigma$$\sigma_{PACE4}^{\alpha xn+2\alpha n}$) calculations.}
\label{fig8}
\end{figure}

For better insight into the onset of $\alpha$-emitting channels, the sum of experimental EFs of $\alpha$-emitting channels, $\Sigma$$\sigma_{exp}^{\alpha xn+2\alpha n}$, is compared with PACE4 ($\Sigma$$\sigma_{PACE4}^{\alpha xn+2\alpha n}$) calculations in Fig.~\ref{fig8}, where the sum is substantially higher by one to two orders of magnitude over the statistical model predictions. This clearly underscores the onset of physical mechanisms such as projectile breakup and partial fragment capture that are not included in the PACE4 formalism and may be attributed to ICF.

\subsubsection{Influence of $\alpha$-emitting channels}
\begin{figure}[]
\centering
\includegraphics[trim=1cm 0.3cm 2.0cm 0.3cm, width=8.4cm]{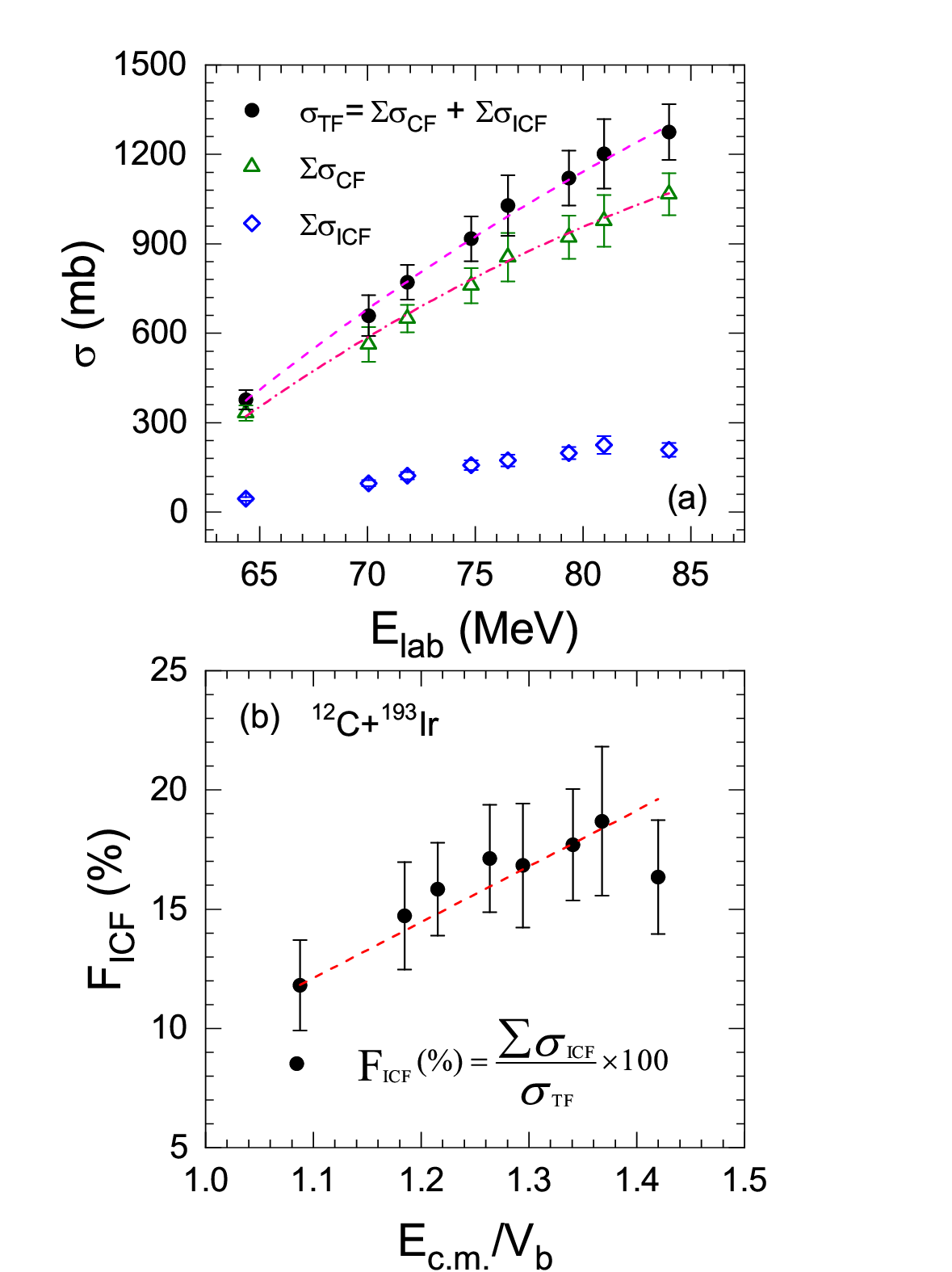}
\caption{$(a)$ The value of $\sigma_{TF}$ along with the $\sigma_{CF}$ and $\sigma_{ICF}$ as a function of incident energy, $(b)$ $F_{ICF}$ as a function of $E_{c.m.}$/$V_{b}$.}
\label{fig9}
\end{figure}

To visualize the influence of ICF in $\alpha$-emitting channels, total fusion cross-section ($\sigma_{TF}$) is plotted in Fig.~\ref{fig9}$(a)$ along with CF ($\sigma_{CF}$) and ICF ($\sigma_{ICF}$) cross-sections. It is essential to note that the activation technique has limitations, particularly for channels with shorter half-lives. Therefore, the $\sigma_{CF}$ value has been corrected by including corresponding unmeasured PACE4 values in CF channels. However, in evaluating $\sigma_{ICF}$, no correction has been made to incorporate the unmeasured ICF channels and is deduced as $\sigma_{ICF} = \sum\sigma_{exp}^{\alpha xn+2\alpha n} - \sum\sigma_{PACE4}^{\alpha xn+2\alpha n}$. Fig.~\ref{fig9}$(a)$ shows the gap between $\sigma_{TF}$ and $\sigma_{CF}$ widens with increasing projectile energy, confirming the progressive enhancement of ICF. To understand how ICF competes with CF, the percentage fraction of ICF, $F_{ICF}$, has been deduced using the data presented in Fig.~\ref{fig9}$(a)$ and plotted against $E_{c.m.}$/$V_{b}$ in Fig.~\ref{fig9}$(b)$. This figure shows that the value of $F_{ICF}$ increases from $\approx12-18\%$, suggesting the breakup probability of $^{12}$C increases with incident energy, likely due to input angular momentum at higher energies, which flattens the fusion pocket in the effective potential. As a result, the projectile breaks up into $\alpha$-clusters to restore the fusion pocket and sustain angular momentum.

\section{\label{sec:level4}Entrance-Channel effect on ICF}
The value of $F_{ICF}$ for various systems has been analyzed in terms of mass asymmetry, Coulomb factor, and neutron skin thickness to understand how entrance-channel parameters affect the onset of ICF.

\subsection{Mass asymmetry ($\mu_{m}$)}
The dependence of entrance-channel mass asymmetry ($\mu_{m}=\frac{A_{T}-A_{P}}{A_{T}+A_{P}}$) on ICF has been investigated for different systems. As proposed by Morgenstern et al. \cite{morgenstern1984}, the systems with higher $\mu_{m}$ contribute significantly to ICF at the same relative velocity, $v_{rel}=[{\frac{2 (E_{c.m.}-V_{b})}{\mu}}]^{1/2}$, where $\mu$ is the reduced mass of the system in $u$, which is converted to $MeV/c^2$, with significant ICF expected above $v_{rel}\approx$ 0.06c (6\% of c). For $^{12}$C+$^{193}$Ir, the $v_{rel}$ lies in the range from $\approx$ 0.03c to 0.06c, clearly demonstrating the onset of ICF at a relatively lower value of $v_{rel}$ $\approx$ 0.03c ($F_{ICF}$ $\approx12\%$). It can be inferred that the ICF begins competing with CF even at slightly above-barrier energies.

\begin{figure}[]
\centering
\includegraphics[trim=1.0cm 0.3cm 1.5cm 0.3cm, width=8cm]{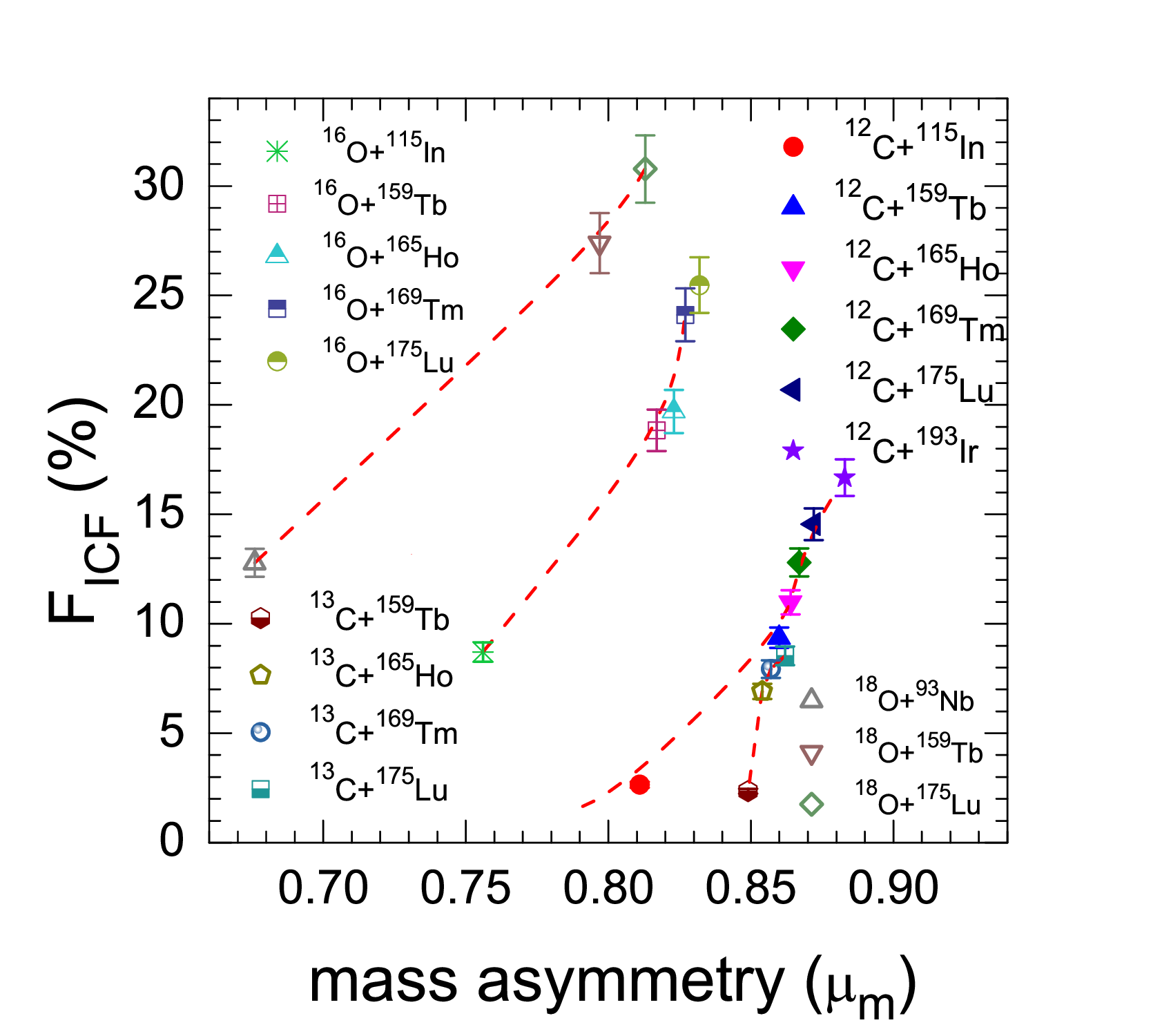}
\caption{$F_{ICF}$ at $v_{rel}$ = 0.053c for various $\alpha$ and non-$\alpha$-cluster projectiles with different targets as a function of $\mu_{m}$.}
\label{fig10}
\end{figure}

Further, Fig.~\ref{fig10} compares $F_{ICF}$ at a constant $v_{rel}$ = 0.053c for projectile-target combinations listed in Table~\ref{tab:table4}. The results show a linear increase in $F_{ICF}$ with $\mu_{m}$ for each projectile ($^{12}$C, $^{13}$C, $^{16}$O, and $^{18}$O). However, $^{16}$O-induced reactions have higher $F_{ICF}$ values as compared to $^{12}$C with the same targets, indicating strong projectile dependence within the mass asymmetry systematics \cite{pps2008}. Based on data presented in Fig.~\ref{fig10}, it can be inferred that the $F_{ICF}$ depends on the projectile structure at energies 5--7 AMeV.

\begin{table}[]
\caption{List of projectile-target combinations displaying the values of their entrance-channel parameters, mass asymmetry ($\mu_{m}$), Coulomb factor ($Z_PZ_T$), and neutron skin thickness ($t_N$) along with $F_{ICF}$ at $v_{rel}$=0.053c.}
\label{tab:table4}
\begin{tabular*}{\columnwidth}{@{\extracolsep{\fill}}llllll@{}}
\hline
System & $\mu_{m}$ & $Z_PZ_T$ & $t_N$ & $F_{ICF}$ (\%) & Ref.\\ \hline
Projectile: $^{12}$C \\
$^{115}$In  & 0.811 & 294 & 0.196  & 2.65 & \cite{mukherjee2006}\\
$^{159}$Tb  & 0.860 & 390 & 0.254  & 9.36  & \cite{yadav_12C_2012}\\ 
$^{165}$Ho  & 0.864 & 402 & 0.264  & 10.98 & \cite{tali2019}\\ 
$^{169}$Tm  & 0.867 & 414 & 0.257  & 12.80 & \cite{chakrabarty2000}\\ 
$^{175}$Lu  & 0.872 & 426 & 0.266  & 14.55 & \cite{harish2017}\\
$^{193}$Ir  & 0.883 & 462 & 0.289  & 17.13 & [This work]\\

Projectile: $^{13}$C \\
$^{159}$Tb  & 0.849 & 390 & 0.254  & 2.35 & \cite{yadav_13C_2012}\\ 
$^{165}$Ho  & 0.854 & 402 & 0.264  & 6.12 & \cite{tali2018}\\ 
$^{169}$Tm  & 0.857 & 414 & 0.257  & 7.93 & \cite{vijay2014}\\ 
$^{175}$Lu  & 0.862 & 426 & 0.266  & 8.54 & \cite{harish2019}\\

Projectile: $^{16}$O \\
$^{115}$In  & 0.756 & 392 & 0.196  & 8.71  & \cite{kamal2014,*kamal2013}\\
$^{159}$Tb  & 0.817 & 520 & 0.254  & 18.84 & \cite{manoj2006,*manoj2003}\\ 
$^{165}$Ho  & 0.823 & 536 & 0.264  & 19.70 & \cite{kkamal2013}\\ 
$^{169}$Tm  & 0.827 & 552 & 0.257  & 24.11 & \cite{pps2008}\\ 
$^{175}$Lu  & 0.832 & 568 & 0.266  & 25.47 & \cite{harish2015}\\

Projectile: $^{18}$O \\
$^{93}$Nb   & 0.676 & 328 & 0.150  & 12.78 & \cite{agarwal2021}\\ 
$^{159}$Tb  & 0.797 & 520 & 0.254  & 27.39 & \cite{yadav2023}\\ 
$^{175}$Lu  & 0.813 & 568 & 0.266  & 30.78 & \cite{harish_thesis}\\ \hline
\end{tabular*}
\end{table}

\subsection{Coulomb factor ($Z_{P}$$Z_{T}$)}
\begin{figure}[]
\centering
\includegraphics[trim=1.0cm 0.3cm 1.5cm 0.3cm, width=8cm]{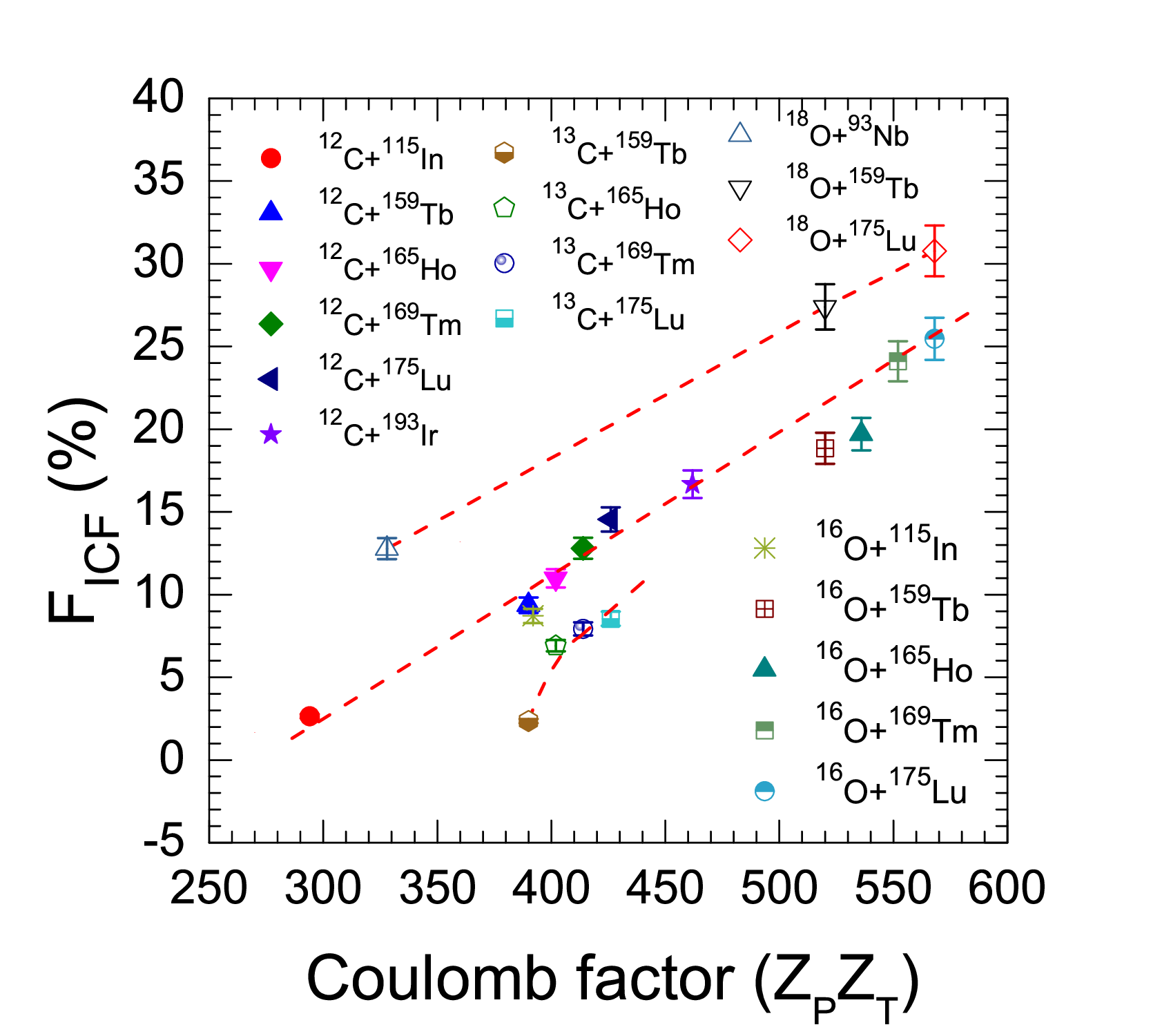}
\caption{$F_{ICF}$ for various systems at a constant $v_{rel}$ = 0.053c as a function of $Z_{P}$$Z_{T}$.}
\label{fig11}
\end{figure}

Fig.~\ref{fig11} displays the ICF strength function in terms of Coulomb factor ($Z_{P}$$Z_{T}$) spanning a range of $\approx$ 294--568 for the systems presented in Table~\ref{tab:table4}. From this figure, $F_{ICF}$ increases linearly with $Z_{P}$$Z_{T}$, indicating a strong influence of the Coulomb force field. As the projectile approaches the target nucleus, the impact of the Coulombic interaction intensifies, thereby increasing the likelihood of projectile breakup followed by the fusion of one or more constituent fragments with the target nuclei. The $\alpha$-cluster projectiles ($^{12}$C and $^{16}$O) exhibit increasing ICF fraction trends and align on the same line, suggesting consistency in their behavior. However, non-$\alpha$-cluster projectiles ($^{13}$C and $^{18}$O) also show increasing trends in ICF, but separately for each projectile, contrary to recent findings \cite{shuaib2016} which demonstrated a linear dependence of ICF on $Z_{P}$$Z_{T}$. Systems with identical $Z_{P}$$Z_{T}$ displayed in Table~\ref{tab:table4} still exhibit different $F_{ICF}$ values, implying that the Coulomb factor alone cannot fully account for the observed ICF behavior. From the data, it may be concluded that the structural characteristics of the projectile also exert a substantial influence on ICF.

\subsection{Target neutron skin thickness ($t_{N}$)}
The neutron skin thickness ($t_{N}$) is a unique property of heavy nuclei due to the imbalance of neutrons and protons. The excess neutrons form a layer that extends beyond the nuclear surface, known as the neutron skin. It is defined as the difference between the matter radius ($R_N$) and the charge radius ($R_Z$) of a nucleus, ($t_N = R_N - R_Z$) \cite{tamii2011,abrahamyan2012}, and can be calculated as,
\begin{equation}
R_{N}=r_0 \left[\frac{2N}{(1-3\epsilon)(1+\delta)}\right]^{1/3}, \\ 
R_{Z}=r_0 \left[\frac{2Z}{(1-3\epsilon)(1-\delta)}\right]^{1/3}.
\label{eq1}
\end{equation} 
The value of $t_{N}$ can be estimated from Eq.~\ref{eq1} \cite{myers1969}.
\begin{equation}
t_{N} = \frac{2}{3} r_0 A^{1/3} (I-\delta).
\label{eq2}
\end{equation} 
where $A$ represents the mass number of the target nucleus, $I$ is a factor defined as $(N - Z)/A$, and $\delta$ is the nuclear density parameter \cite{myers1969}. 

\begin{figure}[]
\centering
\includegraphics[trim=1.0cm 0.3cm 1.5cm 0.3cm, width=8cm]{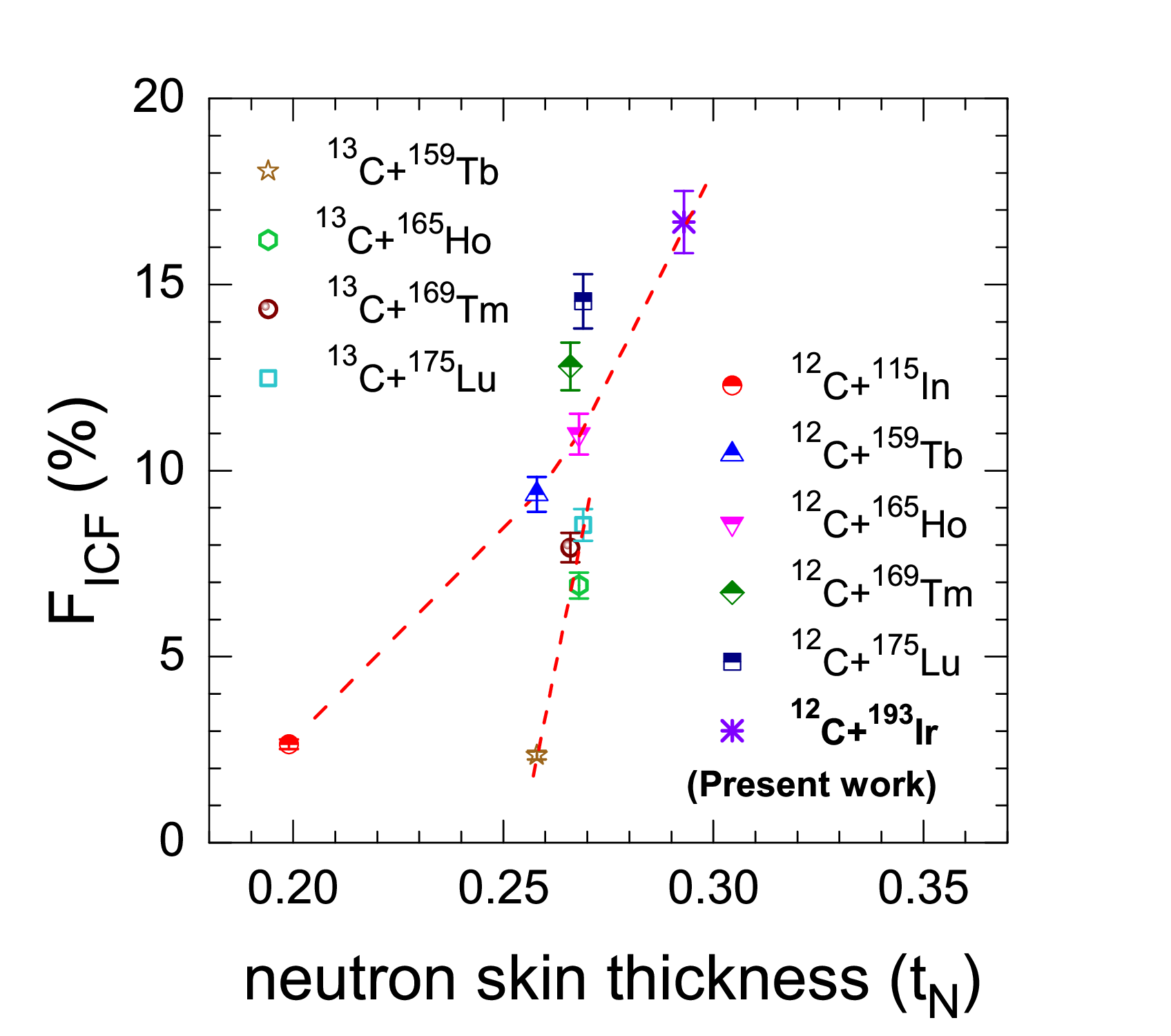}
\caption{The variation of $F_{ICF}$ with neutron skin thickness ($t_N$) at $v_{rel}$ = 0.053c.}
\label{fig12}
\end{figure}

The presence of a neutron skin slightly lowers the Coulomb barrier through screening, thereby enhancing the attractive nuclear potential. To illustrate the influence of neutron skin thickness on ICF, the percentage fraction of ICF is plotted against the neutron skin thickness in Fig.~\ref{fig12}. As evident in the figure, the probability of ICF increases with $t_{N}$ for individual projectiles. This trend suggests a higher likelihood of ICF for neutron-rich, heavy-mass target nuclei. Notably, the $^{12}$C projectile exhibits a higher probability of ICF than the $^{13}$C projectile for the same targets and neutron skin thickness.

\subsection{Projectile structure dependence of ICF}
For a better understanding of the projectile structure dependence of ICF, the radii of $^{12}$C and $^{13}$C have been calculated using the standard relation $R = R_0 A^{1/3}$ (keeping $R_0$ = 1.2 fm), yielding $\approx$ 2.747 fm and 2.821 fm, respectively. No direct correlation between projectile size and $F_{ICF}$ was observed. The $^{13}$C projectile, having one excess neutron, is slightly more weakly bound (binding energy per nucleon 7.469 MeV) than $^{12}$C (7.680 MeV), suggesting a higher breakup probability.

However, from Fig.~\ref{fig10}, ~\ref{fig11} and ~\ref{fig12}, the $F_{ICF}$ values for the $^{13}$C-induced reactions seems to be less than that of the $^{12}$C with the same targets. To understand this difference, it is crucial to consider the characteristics of the projectile nucleus. Projectile $Q_\alpha$ values (-7.37 MeV for $^{12}$C, -10.65 MeV for $^{13}$C, -7.16 MeV for $^{16}$O, and -6.23 MeV for $^{18}$O) reveal that systems with less negative $Q_\alpha$ values exhibit larger $F_{ICF}$. This indicates that weaker $\alpha$-separation energies favor the breakup and subsequent fusion of fragments. A series of experiments have been performed using $^{12}$C projectile, with $^{103}$Rh \cite{bindu1999,buthelezi2004}, $^{115}$In \cite{mukherjee2006}, $^{128}$Te \cite{manoj2003}, $^{159}$Tb \cite{jashwal2023}, $^{160}$Gd \cite{siwek_1979}, $^{165}$Ho \cite{tali2019,tali2018,ojha2021}, $^{169}$Tm \cite{sahoo2019,chakrabarty2000}, $^{175}$Lu \cite{harish2017,harish2019}, $^{181}$Ta \cite{babu2003_ef,*babu2004_rrd,pavneet_181_2022,*pavneet_197_2022}, $^{197}$Au \cite{pavneet_197_2022,vergani1993}, $^{208}$Pb \cite{kalita2011} targets, in which ICF has been attributed to the low $\alpha$ separation energy of $^{12}$C projectile into $^{8}$Be+$\alpha$, and/or $\alpha$+$\alpha$+$\alpha$ fragment configurations \cite{valdes2023}. Overall, the results highlight a strong dependence of ICF on projectile structure, distinguishing between $\alpha$-cluster and non-$\alpha$-cluster nuclei and establishing $Q_\alpha$ as a key entrance-channel parameter, thereby providing a deeper understanding of the impact of projectile structure on ICF dynamics.

\subsection{Onset of ICF below $\ell_{crit}$}
Input angular momentum ($\ell$) is also a sensitive entrance-channel parameter reported to be responsible for the low-energy CF. To gain deeper insight into the fusion angular momentum distribution, the maximum angular momentum ($\ell_{max}$) and the critical angular momentum ($\ell_{crit}$) values have been computed for the $^{12}$C+$^{193}$Ir system. The $\ell_{crit}$ is the limiting angular momentum for fusion to occur and may be calculated from
\begin{equation}
   \ell_{crit}^2 = \frac{\mu (C_1+C_2)^3}{\hbar^2} \left[4\pi \gamma \frac{C_1 C_2}{C_1 + C_2} -  \frac{Z_P Z_T e^2}{(C_1 + C_2)^2}\right],
    \label{eq3}
\end{equation} 
where $\mu$ is the reduced mass of the system, $\gamma$ is the surface tension coefficient given as $\gamma = 0.95(1-1.78 I^2)\, \text{MeV/fm}^2,  \quad I= \frac{N-Z}{A}$, and $Z_P, Z_T$ are the atomic numbers of the projectile and target nuclei, and $C_1, C_2$ are the half-density radii of the projectile and target nuclei, given as
$C = R\left(1 - \frac{b^2}{R^2}\right), \quad b = 1 \,\text{fm}$, where $R$ is the normal radius estimated as $R = 1.28A^{1/3} - 0.76 + 0.8A^{-1/3} \,\text{fm}$.

The deduced value of $\ell_{crit}$ using Eq.~\ref{eq3} is $\approx$ 43$\hbar$ for $^{12}$C+$^{193}$Ir system. However, if we replace the half-density radii ($C$) with the sharp radius $R = 1.11 A^{1/3}\,\text{fm}$ \cite{wilczynski1973}, then the estimated value of $\ell_{crit}$ is $\approx$ 51$\hbar$ for $^{12}$C+$^{193}$Ir system. This suggests a range of $\ell_{crit}$ from 43$\hbar$ to 51$\hbar$ for various peripheral processes, including ICF and nucleon or cluster transfer, in $^{12}$C-induced reaction on $^{193}$Ir. The $\ell_{max}$ estimated for $^{12}$C+$^{193}$Ir system at all studied energies, using the CCFULL code \cite{hagino1999,*hagino2003} employing Akyuz-Winther parametrization (potential depth (\( V_0 = 59.042 \) MeV), radius (\( r_0 = 1.178 \) fm), and diffuseness (\( a_0 = 0.646 \) fm)) is found to be $\approx$ \( 36\hbar \) and \( 15\hbar \) at \( E_{\text{c.m.}} = 79.1 \) MeV and 60.6 MeV, respectively. The $\ell_{max}$ = 36$\hbar$ at 79 MeV $<$ $\ell_{crit}$=43--51$\hbar$ suggests diffused $\ell$-window, consistent with coupled-channel calculations showing 10--20\% ICF from $\ell < \ell_{crit}$ \cite{hagino2003}.

\begin{figure}[]
\centering
\includegraphics[trim=1.0cm 0.3cm 1.5cm 0.3cm, width=8cm]{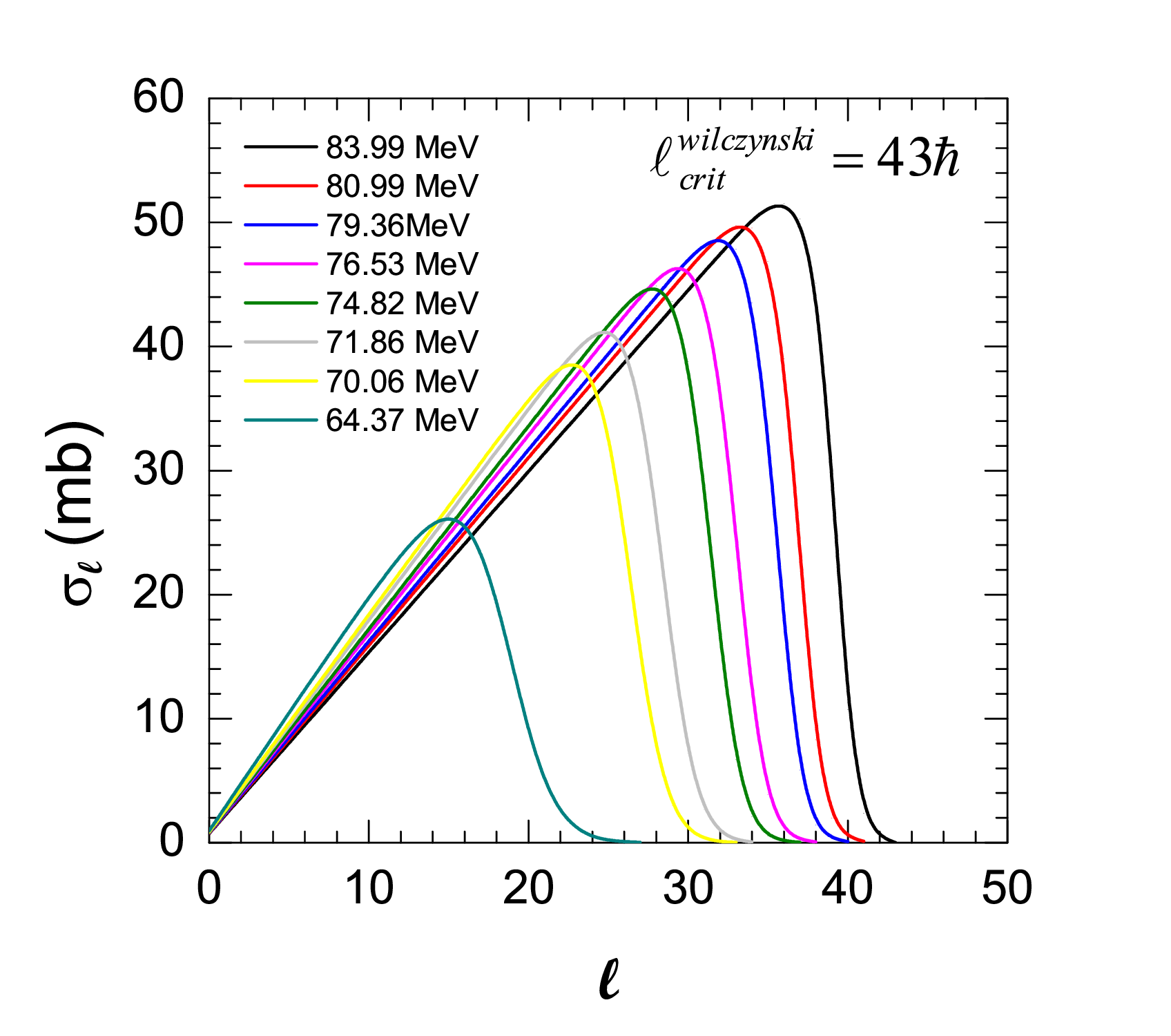}
\caption{fusion $\ell$ distributions estimated using the code CCFULL to understand the population of $\ell$-bins at each studied lab energy.}
\label{fig13}
\end{figure}

It can be estimated from Fig.~\ref{fig13} that the value of $\ell_{max} < \ell_{crit}$ (43\( \hbar \)) for energies 64--84 MeV. This figure demonstrates that even at the highest incident beam energy, $\ell < \ell_{crit}$, indicating that a significant portion of \( \ell \)-bins below \( \ell_{crit} \) may contribute to the ICF process. The present findings highlight a diffuse boundary for \( \ell \)-values, in contrast to the predictions of the sharp-cut-off model, suggesting that angular momentum contributions extend into the region near the barrier.

\section{\label{sec:level5}Fusion Suppression due to projectile breakup}
It is better to use a standard reduction procedure suggested by Canto et al. \cite{canto2009,*canto_2009,*canto2024} to obtain a systematic behavior of fusion cross-section data by eliminating the geometrical and static effects due to the potential acting between the interacting nuclei.

\subsection{Universal Fusion Function (UFF)}
\begin{figure}[]
\centering
\includegraphics[trim=1.0cm 0.3cm 1.5cm 0.3cm, width=8cm]{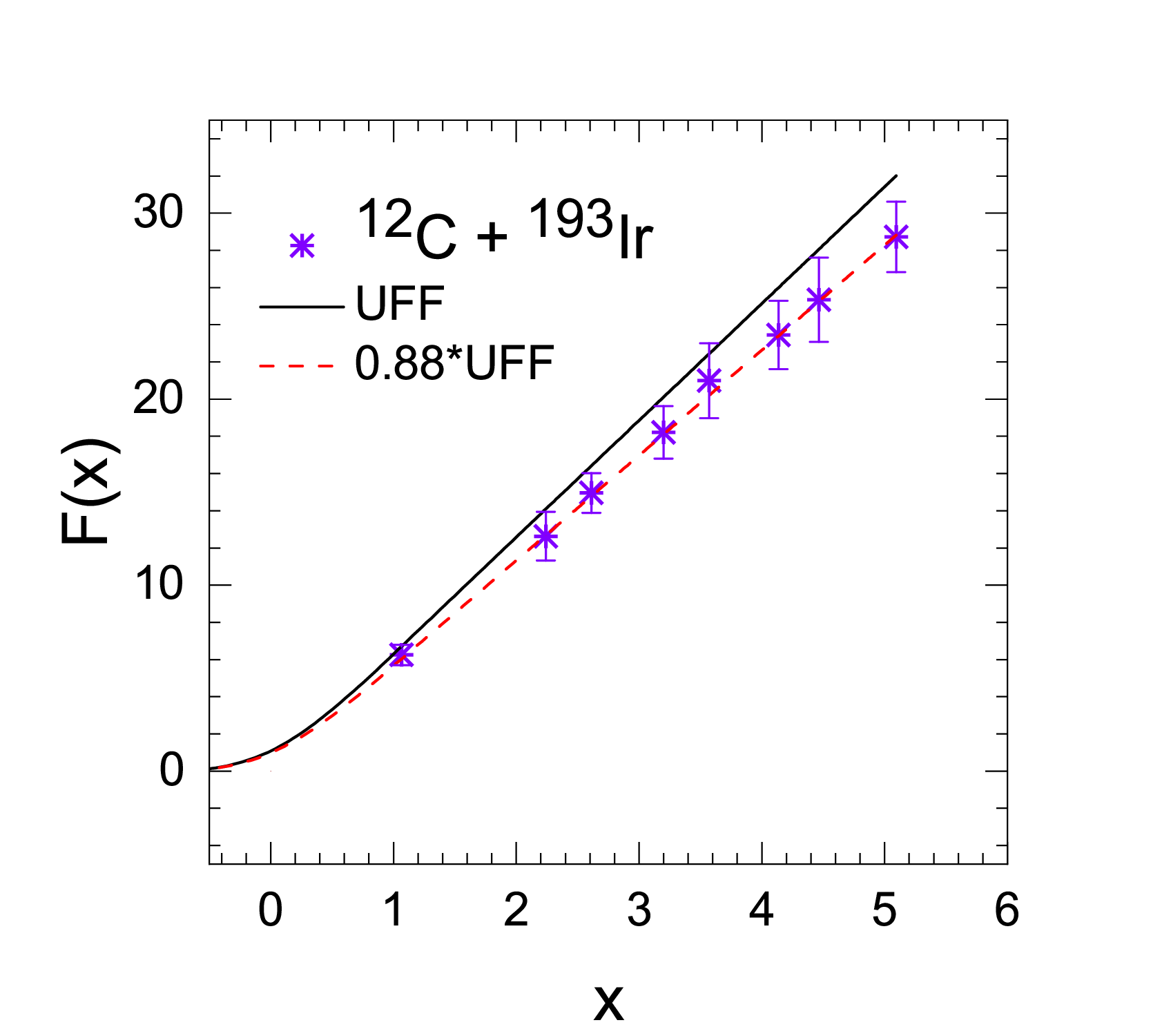}
\caption{The fusion function for $^{12}$C+$^{193}$Ir system indicating a suppression of $\approx12\%$ in CF compared to UFF.}
\label{fig14}
\end{figure}

The experimental fusion cross-sections ($\sigma_{F}$) have been transformed into a dimensionless function $F(x)$, and the incident energy into a dimensionless variable $x$ \cite{canto2009} as,
\begin{equation}
   E_{c.m.} \rightarrow x = \frac{E_{c.m.} - V_{b}}{\hbar\omega},
   \sigma_{F} \rightarrow F(x) = \frac{2 E_{c.m.}}{\hbar\omega R_{b}^2} \sigma_{F},
    \label{eq4}
\end{equation}
where $E_{c.m.}$ is the center-of-mass energy, $\sigma_{F}$ is the fusion cross-section, and $\hbar\omega$, $R_{b}$, and $V_{b}$ represent the barrier curvature, barrier radius, and barrier height, respectively. The values of barrier parameters, $\hbar\omega$ = 4.590 MeV, $R_{b}$ = 11.313 fm, and $V_{b}$ = 55.709 MeV are sourced from the NRV code \cite{nrv}. This method is inspired by Wong's formula for fusion cross-section \cite{wong1973},
\begin{equation}
   \sigma_{F}^{W} (E_{c.m.}) = \frac{R_{b}^2 \hbar\omega}{2 E_{c.m.}} \ln{\left[1+\exp\left(\frac{2 \pi (E_{c.m.} - V_{b})}{\hbar\omega}\right)\right]}.
    \label{eq5}
\end{equation}
The Universal Fusion Function (UFF), $F_{0}(x)=\ln{[1+\exp(2 \pi x)]}$, is constant across different systems. Any deviation from the UFF may be attributed to the effects of projectile breakup upon CF cross-sections \cite{wang2014,*wang2016}. The UFF for the $^{12}$C+$^{193}$Ir system is shown in Fig.~\ref{fig14}, with the dotted line representing the UFF multiplied by the suppression factor, $F_{B.U.}=\frac{F(x)}{F_{0}(x)}$, which is 0.88 times the UFF, determined by fitting the experimental fusion function data in Fig.~\ref{fig14}. 

Wang's empirical formula may be used to correlate the suppression with breakup threshold energy ($E_{B.U.}$) \cite{wang2014,*wang2016},
\begin{equation}
   \log_{10}(1-F_{B.U.}) = -a \exp\left(\frac{-b}{E_{B.U.}}\right) -c E_{B.U.},
    \label{eq6}
\end{equation}
where $a$, $b$, and $c$ are the fitting parameters, and their values by Wang et al. are 0.33, 0.29 MeV, and 0.087 MeV${^{-1}}$.

\begin{figure}[]
\centering
\includegraphics[trim=1.0cm 0.3cm 1.5cm 0.3cm, width=8cm]{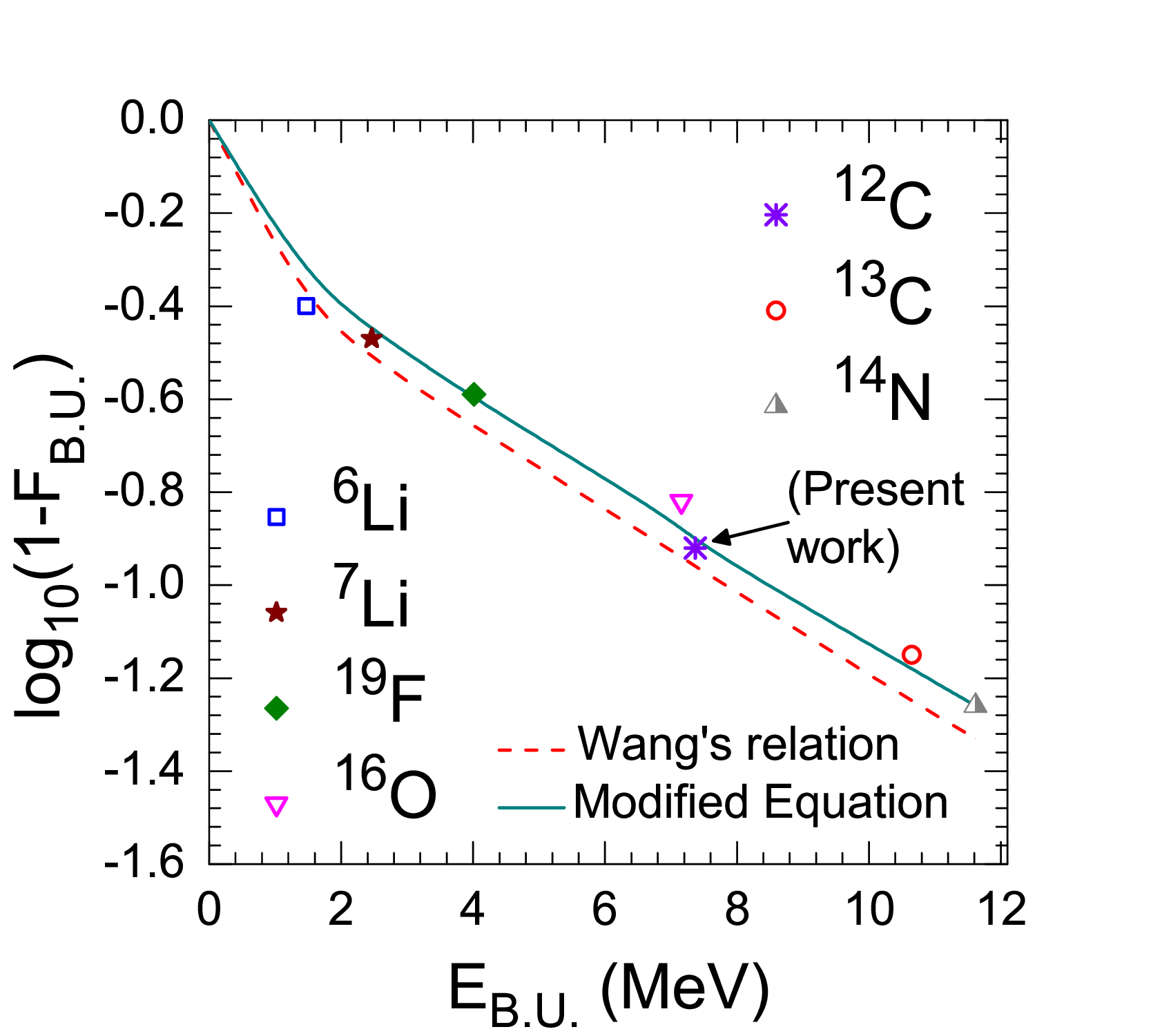}
\caption{The $F_{B.U.}$ for different projectiles with their $E_{B.U.}$. The dotted line corresponds to Wang's equation, while the solid line represents modified parameters.}
\label{fig15}
\end{figure}

In Fig.~\ref{fig15}, the variation of $F_{B.U.}$ for different projectiles is plotted as a function of their $E_{B.U.}$. The dotted line represents the suppression trend as described by Wang's parameters. However, the experimental data do not agree with this. Therefore, the constants ($a$, $b$, and $c$) in Eq.~\ref{eq6} have been adjusted to 0.28, 0.39 MeV, and 0.085 MeV${^{-1}}$. As shown in Fig.~\ref{fig15}, after implementing the adjusted constants, the experimental data for different projectiles agree reasonably well with the modified equation, unlike those obtained using Wang's parameters. The adjusted parameters preserve the overall functional form of Wang's empirical relation while improving its accuracy in the intermediate breakup-threshold region. Further, the $\log_{10}(1-F_{B.U.})$ values from modified parameters, fitted to the experimental data for various systems, are listed in Table~\ref{tab:table5}. This correlation indicates that $E_{B.U.}$ strongly governs projectile breakup and the degree of fusion suppression and, thus, the fusion cross-section.

\begin{table}[h!]
\caption{\label{tab:table5}The value of $F_{B.U.}$ for various projectiles from Wang formulation \cite{wang2014,*wang2016} and modified equation.}
\begin{tabular*}{\columnwidth}{@{\extracolsep{\fill}}llllll@{}}
\hline
 Projectile & $E_{B.U.}$ & Wang's & Modified & $F_{B.U.}$ & $\log_{10}$ \\
  &   (MeV) & relation & equation & (fit) & $(1-F_{B.U.})$ \\ \hline
  $^{6}$Li  & 1.473 & -0.40 & -0.34 & 0.60 & -0.40  \\ 
  $^{7}$Li  & 2.467  & -0.51 &-0.45 & 0.66 & -0.47 \\ 
  $^{12}$C  & 7.367  & -0.96 &-0.89 & 0.88 & -0.92 \\ 
  $^{13}$C  & 10.648 & -1.25 &-1.18 & 0.93 & -1.15 \\
  $^{14}$N  & 11.62 & -1.33 &-1.26 & 0.945 & -1.26 \\ 
  $^{16}$O & 7.161 & -0.94 &-0.87 & 0.85 & -0.82 \\ 
  $^{19}$F & 4.014 & -0.66 &-0.60 & 0.74 & -0.59 \\ \hline
\end{tabular*}
\end{table}

\subsection{Improved Fusion Function (IFF)}
Canto \textit{et al.} \cite{canto2024} improved Wong's formula by replacing the parabolic approximation of the Coulomb barrier with the $\ell$-dependent potential at an effective partial wave. The transformations yield the Classical Fusion Line (CFL) as a benchmark for comparing fusion data. 
\begin{equation}
    E_{c.m.} \rightarrow y = 1 - \frac{V_{b}}{E_{c.m.}}; \
   \sigma_{F} \rightarrow \Bar{G}^{exp}(y) = \frac{\sigma_{F}^{exp}}{\pi f_{app}(y) R_b^2},
    \label{eq7}
\end{equation}
where $V_b$ is the barrier height and $\Bar{R}$ is the effective barrier radius. The empirical function $f_{app}(y)$ is approximated as, $f_{app}(y) \approx f_{R}(y) = 1 - 0.14 y - 0.14 y^2$.

\begin{figure}[]
\centering
\includegraphics[trim=1.0cm 0.3cm 1.5cm 0.3cm, width=8cm]{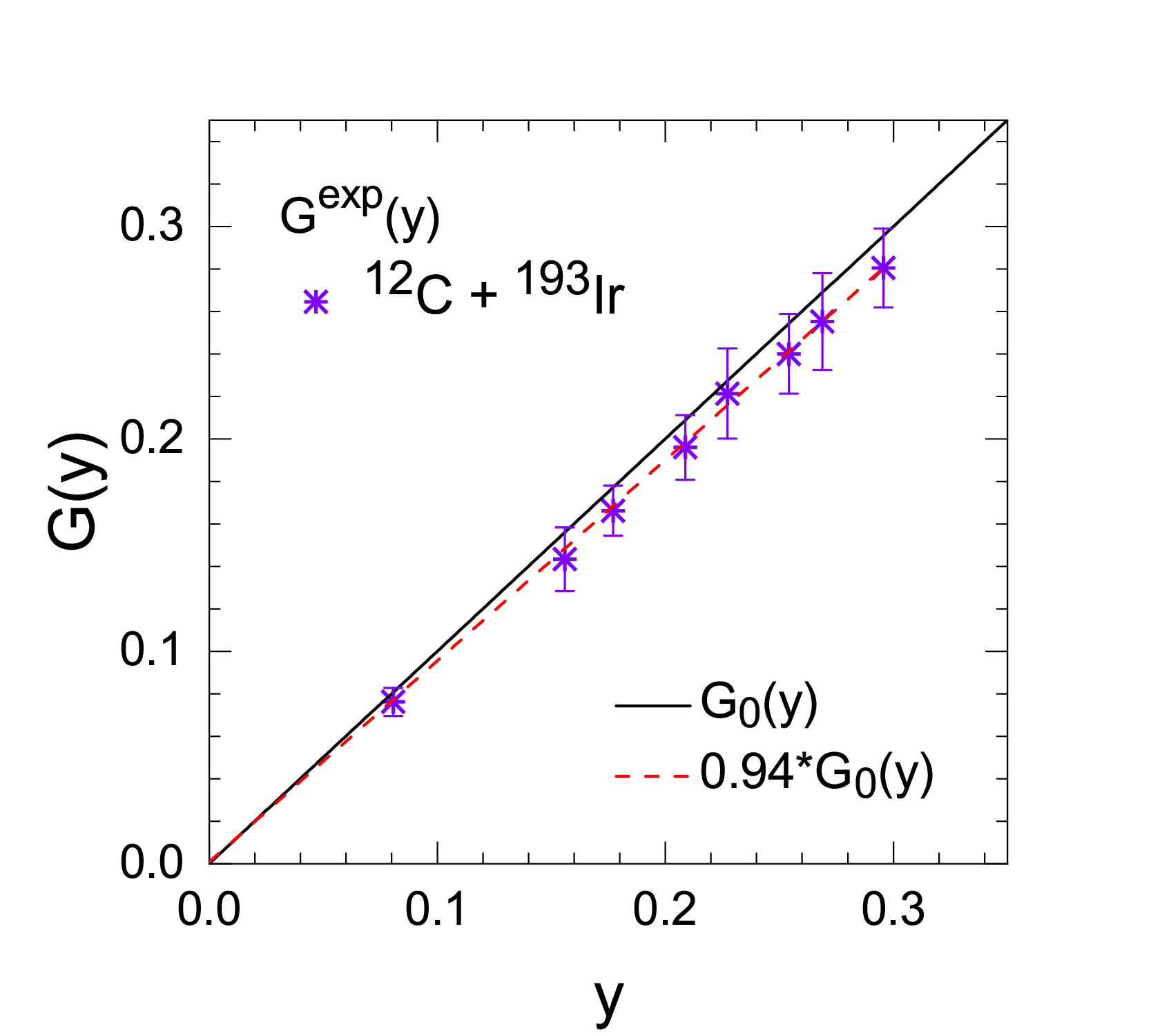}
\caption{Experimentally obtained IFF for $^{12}$C+$^{193}$Ir system. The solid line denotes the classical fusion line. The lines and symbols are self-explanatory.}
\label{fig16}
\end{figure}

The improved classical cross-section yields another universal function, denoted $G_0(y)$.
\begin{equation}
   E_{c.m.} \rightarrow y = 1 - \frac{V_{b}}{E_{c.m.}}, \quad
   \bar{\sigma}_{cl} \rightarrow G_0 = \frac{\bar{\sigma}_{cl}}{\pi \Bar{R}^2}
    \label{eq8}.
\end{equation}
From this equation, we obtain the CFL, $G_0(y) = y.$ The Improved Fusion Function (IFF) for $^{12}$C+$^{193}$Ir system is presented in Fig.~\ref{fig16}, where the reduced fusion data is very close to the CFL with a suppression of 6$\%$ within the error bars. The IFF analysis indicates that the observed difference in UFF mainly arises from the $\ell$-dependence of the barrier parameters, which is incorporated in IFF by replacing $R_b^2$ by ${\Bar{R}^2} = R_b^2 \times f_R(y)$, improving the description at higher energies without altering the basic interpretation of ICF.

\section{\label{sec:level5}Summary and conclusion}
In summary, channel-by-channel EFs of different ERs have been measured in $^{12}$C+$^{193}$Ir system at $\approx$ 5--7 AMeV and analyzed in the framework of PACE4. The EFs of $xn$ and $pxn$ channels have been well reproduced by PACE4 calculations with input level density parameter $a$ = A/13 MeV${^{-1}}$, indicating the population of these residues by CF. However, the $pxn$ channels are populated via precursor decay via their higher-charge isobars. Additionally, the EFs of $\alpha$-emitting channels display significant enhancement over PACE4 calculations, indicating the onset of ICF. For a better insight into the onset and influence of ICF, the $F_{ICF}$ has been analyzed in the framework of entrance-channel mass asymmetry, Coulomb effect, and target neutron skin thickness. It has been found that the probability of ICF increases from 12--18\% with increasing beam energy, indicating its sensitivity to the projectile energy. The $F_{ICF}$ value in light of $\mu_{m}$ shows a strong dependence on the projectile type. Low-energy ICF data for different systems, with $Z_{P}$$Z_{T}$ (294--568) and $t_{N}$, indicate that breakup probability depends on both projectile structure and $Q_\alpha$ value. To analyze ICF data, no single parameter thoroughly explains ICF dynamics; projectile structure, $Q_\alpha$, and entrance-channel parameters must be considered. The sharp cut-off model significantly underestimates the ICF cross-sections, suggesting that a notable part of ICF may originate from collision trajectories with $\ell \leq\ell_{crit}$. This discrepancy indicates that the fusion window assumed by the model may require refinement, particularly at 5--7 AMeV. Further, the fusion function derived from CF cross-sections is found to be $\approx$12$\%$ lower than UFF, attributed to prompt projectile breakup. Moreover, Wang's empirical relation \cite{wang2014,*wang2016} confirms that the projectile breakup is influenced by its breakup threshold energy. The IFF approach was adopted by incorporating $\ell$-dependent barrier parameters, and IFF-reduced fusion data showed only $6\%$ suppression relative to CFL, indicating that the earlier UFF-based analysis likely overestimated ICF. The present study thus provides quantitative constraints on low-energy breakup fusion dynamics in an $\alpha$-cluster projectile system. Future measurements with neighboring projectiles, such as $^{16}$O+$^{193}$Ir, would help isolate pure Coulomb effects from $Q_\alpha$-driven breakup mechanisms. These results contribute to improving near-barrier reaction models and nuclear data, both of which are important for nuclear structure studies and astrophysical reaction-rate calculations.

\begin{acknowledgements}
The authors acknowledge the Inter-University Accelerator Centre (IUAC), New Delhi, for facilities to carry out this experiment, and Dr. Rakesh Kumar (deceased) for contributing to this work. One of the authors, Amanjot, thanks the Department of Science $\&$ Technology (DST), Govt. of India, for the INSPIRE Doctoral Fellowship (DST/INSPIRE Fellowship/2019/IF190067). 
\end{acknowledgements}

\bibliography{bib}

\end{document}